\documentclass[twocolumn,showpacs,preprintnumbers,amsmath,amssymb,aps,prc]{revtex4-2}
	
\usepackage{xcolor}
\usepackage{verbatim}
\usepackage{graphicx}
\usepackage{dcolumn}
\usepackage{bm}
\usepackage{lineno}
\usepackage{multirow}

\usepackage[colorlinks=true, allcolors=blue]{hyperref}
\usepackage[caption=false]{subfig}

\begin{document}
\title{ Modeling Backward-Angle ($u$-channel) Virtual Compton Scattering at an Electron-Ion Collider}
\author{Zachary Sweger}
\affiliation{Department of Physics and Astronomy, University of California, Davis, California 95616, USA}
\author{Spencer R. Klein}
\affiliation{Lawrence Berkeley National Laboratory, Berkeley, California 94720, USA}
\author{Yuanjing Ji}
\affiliation{Lawrence Berkeley National Laboratory, Berkeley, California 94720, USA}
\author{Minjung Kim}
\affiliation{Department of Physics, University of California, Berkeley, California 94720, USA}
\author{Saeahram Yoo}
\affiliation{Department of Physics and Astronomy, University of California, Davis, California 95616, USA}
\author{Ziyuan Zeng}
\affiliation{Department of Physics and Astronomy, University of California, Davis, California 95616, USA}
\author{Daniel Cebra}
\affiliation{Department of Physics and Astronomy, University of California, Davis, California 95616, USA}
\author{Xin Dong}
\affiliation{Lawrence Berkeley National Laboratory, Berkeley, California 94720, USA}
\date{\today}

\begin{abstract}

High-energy backward ($u$-channel) reactions can involve very large momentum transfers to the target baryons, shifting them by many units of rapidity.   These reactions are difficult to understand in conventional models in which baryon number is carried by the valence quarks. Backward Compton scattering is an especially attractive experimental target, because of its simple final state. There is currently limited data on this process, and that data is at low center-of-mass energies. In this paper, we examine the prospects for studying backward Compton scattering at the future Electron-Ion Collider (EIC). We model the cross-section and kinematics using the limited data on backward Compton scattering and backward meson production, and then simulate Compton scattering at EIC energies, in a simple model of the ePIC detector.  Generally, the proton is scattered toward mid-rapidity, while the produced photon is in the far-forward region, visible in a Zero Degree Calorimeter (ZDC). We show that the background from backward $\pi^0$ production can be rejected using a high-resolution, well-segmented ZDC.

\end{abstract}

\maketitle

\section{Introduction}

\begin{figure}[tp]

\subfloat[]{%
\includegraphics[clip,width=0.5\columnwidth \label{fig:COMFrame}]{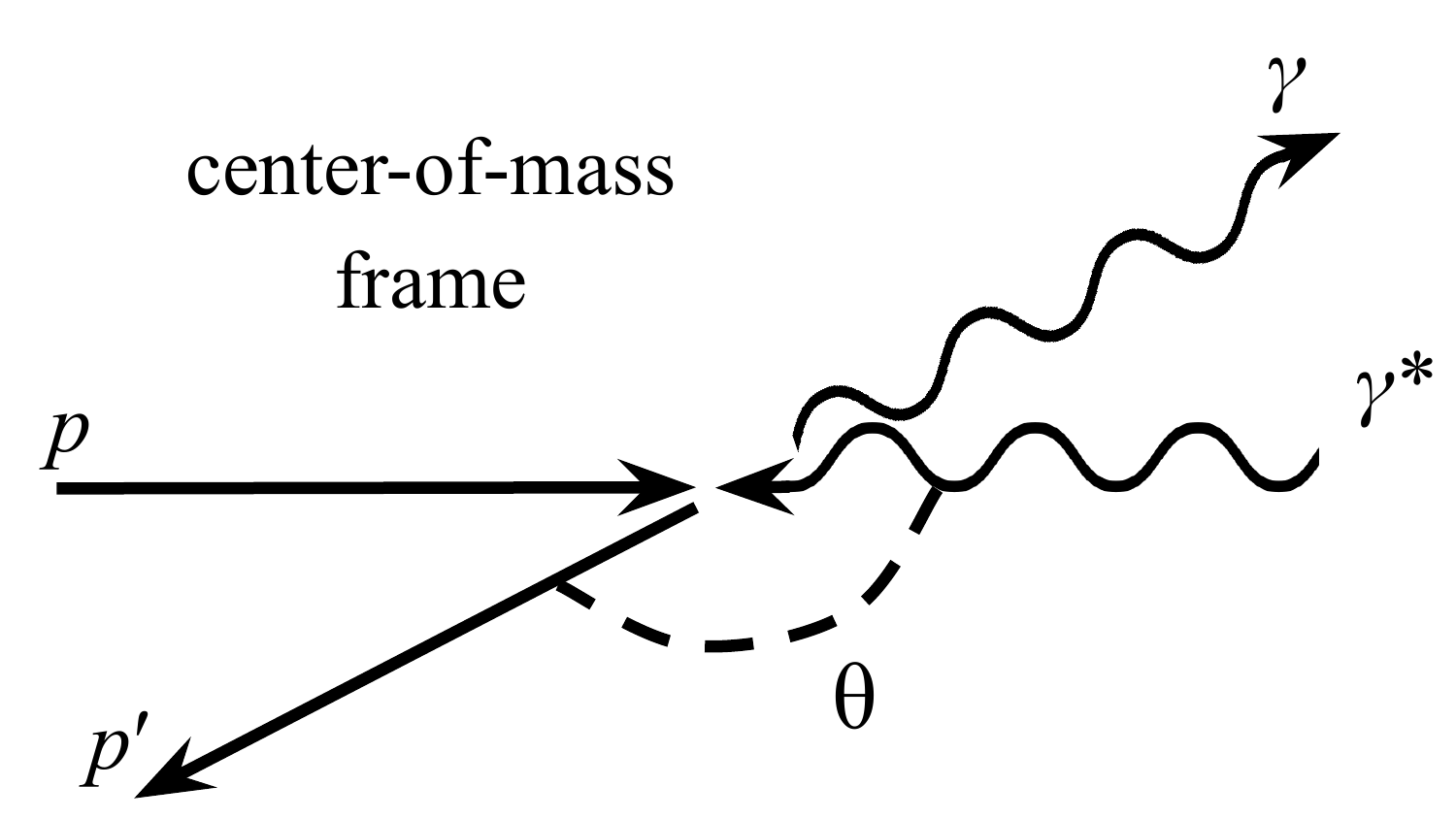}%
}%

\subfloat[]{%
\includegraphics[clip,width=0.7\columnwidth \label{fig:Mandelstam}]{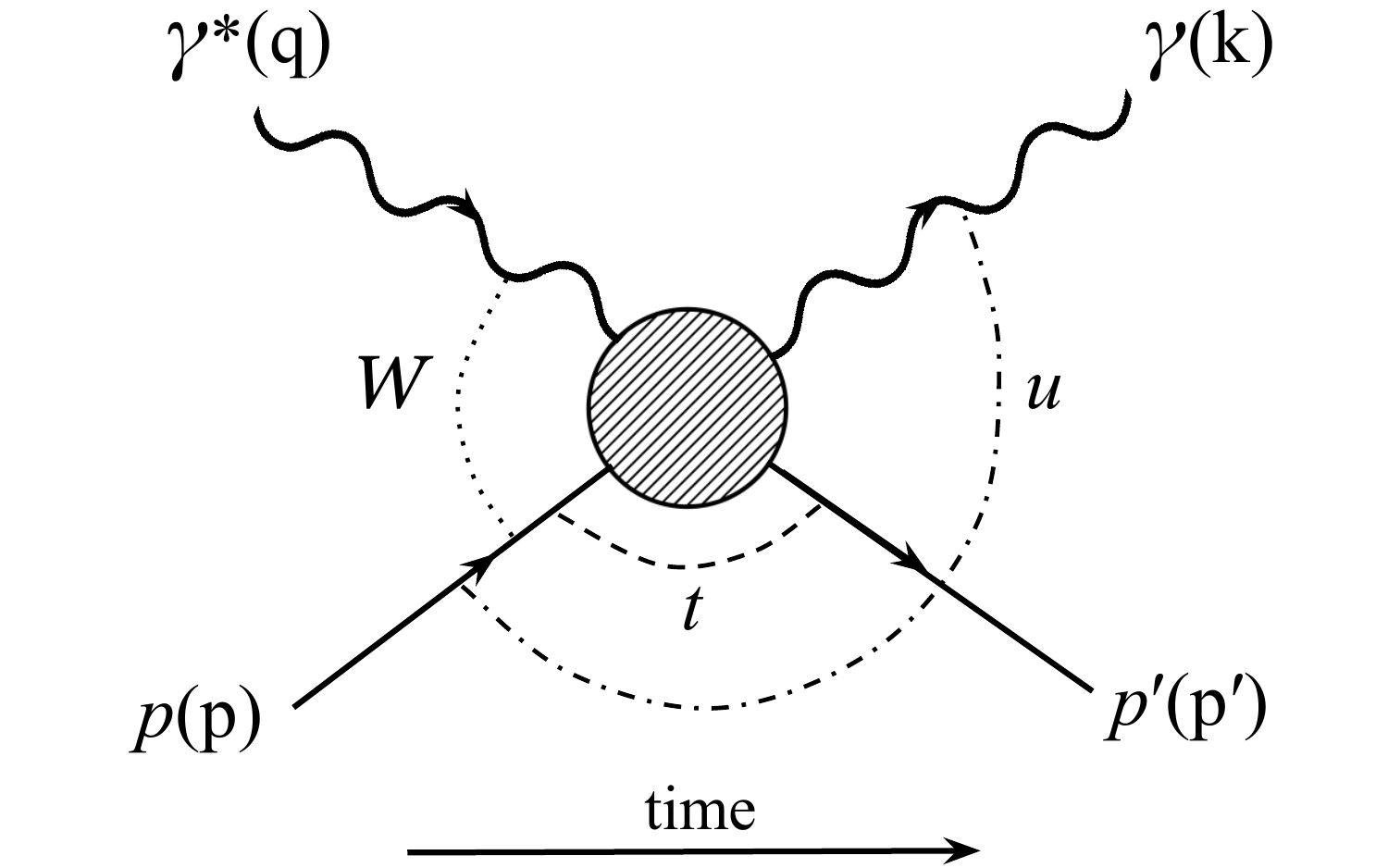}%
}%

\caption{(a) VCS and the proton/photon scattering angle as seen from the center-of-mass frame. Contributions to the cross section from $u$-channel exchange are expected to dominate at $\theta\sim 180^{\circ}$. (b) Initial and final states in VCS, their four-momenta, and associated kinematic variables $W$, $u$, and $t$.}
\end{figure}

Backward ($u$-channel) Compton scattering (CS) occurs when a photon scatters backwards from a proton, with a large momentum transfer between the two as shown in Fig.~\ref{fig:COMFrame}.  This is in stark contrast to the more common $t$-channel process which dominates the CS cross section. In $t$-channel (forward/small-angle) Compton scattering, the momentum transfer between the photon and proton is small, as is the scattering angle, e.g. $\theta\approx0$, so $|t|\approx0$. $u$-channel CS has a near-maximal momentum transfer $|t|$, and small $|u|$ with $\theta\approx180^\circ$. 

If the initial-state photon is a virtual photon, Compton scattering is referred to as virtual Compton scattering (VCS) or deeply-virtual Compton scattering (DVCS) depending on the photon's invariant mass squared ($-Q^2$).
DVCS is considered a ``golden channel" of the future U.S. Electron-Ion Collider because $t$-channel DVCS provides access to proton Generalized Parton Distributions (GPDs) at non-zero skewedness~\cite{BURKARDT_2003}. 
Much attention has been paid to $t$-channel Compton scattering due to its dominance of the total cross section and straightforward interpretation in terms of GPDs.  

Forward DVCS is used for proton tomography~\cite{2019188} because the transverse component of the Mandelstam $t$ is conjugate to the distribution of partons in the transverse plane~\cite{Diehl:2002he}, i.e. as a function of impact parameter. 
The majority of DVCS measurements to-date have been collected at low $|t|$ primarily from experiments at Jefferson Lab~\cite{Defurne:2017paw,2019188} and DESY~\cite{H1:2007vrx,ZEUS:2003pwh}. These low-$t$ measurements map the proton at large impact parameters, but there is little to constrain parton distributions at small radii. For this reason, the EIC White Paper~\cite{WhitePaper} stresses the need for DVCS measurements up to large momentum transfers.

Recent theoretical work on baryon-to-photon/meson transition distribution amplitudes (TDAs) proposes a $u$-channel factorization scheme similar to the $t$-channel factorization with an impact-parameter interpretation of backward amplitudes~\cite{Pire:2021hbl}. In this view, the TDAs encode information about the transverse distribution of di-quark and tri-quark clusters within the proton. Furthermore, backward DVCS and other $u$-channel processes may play a role in baryon stopping in heavy-ion collisions, in which nucleons undergo large momentum transfers and are detected near midrapidity~\cite{PhysRevC.106.015204}.

VCS analyses at the EIC should attempt to measure the magnitude of the $u$-channel contribution to the VCS cross section, and how it scales with $Q^2$, $W$, and $t$. After transforming the cross section from transverse momentum to impact-parameter space, these measurements may allow the EIC to map those partonic constituents that contribute to reactions involving baryon-number transfer.

Moreover, without knowing if a backward VCS peak exists, the magnitude of $u$-channel contributions to forward DVCS cross section measurements at the EIC are unknown. Toward low (threshold) $\gamma^* p$ collision energies, the difference between the $|t|$ values corresponding to forward and backward scattering becomes small. As a result, it is difficult to isolate the $t$-channel contribution at threshold, as the $u$-channel mechanism may contribute an unknown amount. Therefore it may benefit DVCS studies to understand the magnitude of the contribution that $u$-channel exchange adds to the cross section. 
    
In this paper, we examine the prospects of measuring backward VCS at the EIC, in the face of a large background from backward $\pi^0$ production. In Sec.~\ref{section:generalkinematics}, we define kinematic variables and provide background information on backward Compton scattering.  Section~\ref{section:themodel} presents a model of backward VCS developed from existing data. Section~\ref{section:backgrounds} discusses backgrounds to the backward VCS signal, including $u$-channel $\pi^0$ production. Details about the simulations that were developed are provided in Sec.~\ref{section:simulations}. Section~\ref{section:detection} discusses the prospects for detecting these simulated events at the EIC, given current detector expectations. This section also demonstrates how the background may be reduced to a few percent of the backward VCS signal.

\section{Kinematics}
\label{section:generalkinematics}
\begin{figure}
  \begin{center}
    \includegraphics[width=0.5\textwidth]{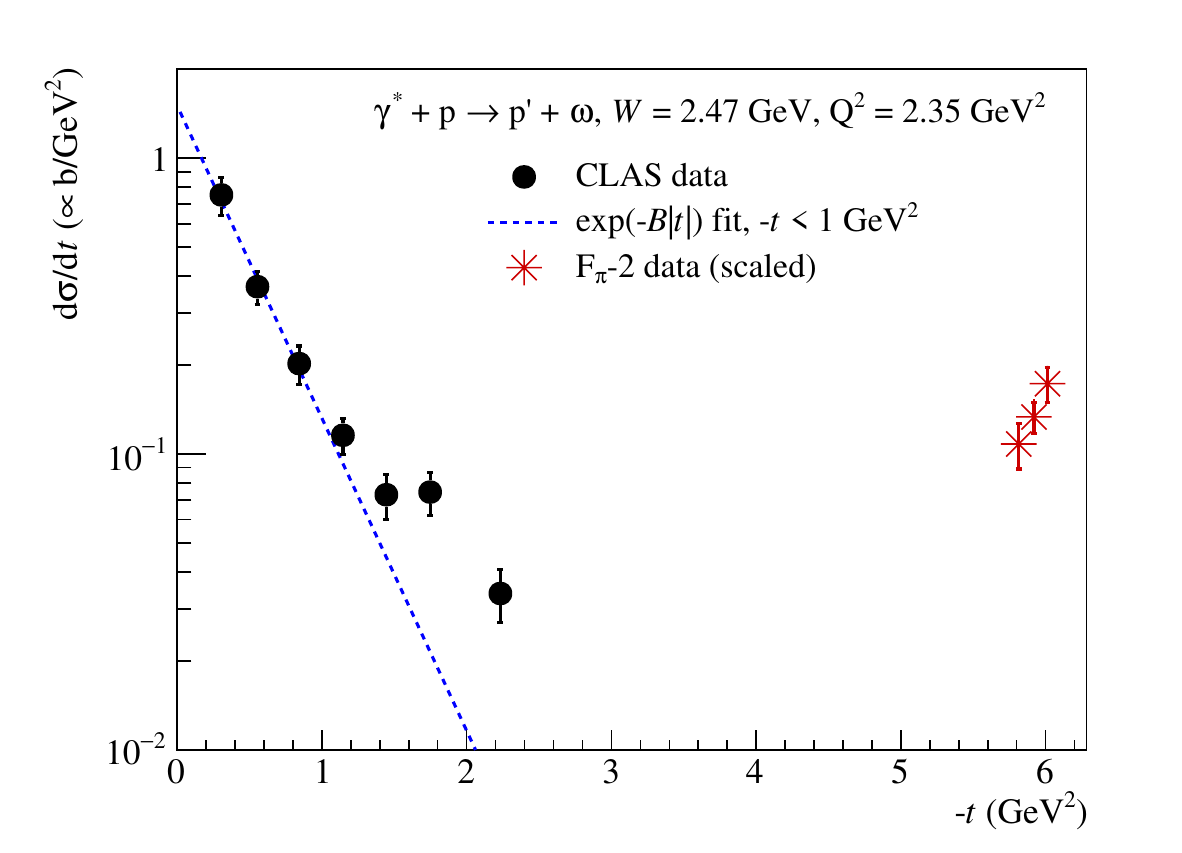}
    \caption{Differential cross section for $\omega$ electroproduction as a function of $-t$ from Ref.~\cite{billspaper,Gayoso:2021rzj}, demonstrating a peak at backward angles (large $|t|$). Forward-scattering data from the CLAS Collaboration~\cite{CLAS_Morand_2005} and backward data from the F$_\pi$-2 experiment~\cite{PhysRevLett.97.192001} are compared. The CLAS data is fit with an exponential (to guide the eye) as in Eq.~\ref{eq:dsigdt} for $-t<1$~GeV$^2$. The F$_\pi$-2 data were collected at $W=2.21$~GeV and $Q^2=2.45$~GeV$^2$. They were scaled according to their kinematics to be comparable to the CLAS data at $W=2.47$~GeV and $Q^2=2.35$~GeV$^2$.}
    \label{fig:backwardspeak}
  \end{center}
\end{figure}
The kinematic variables used to describe DVCS and electroproduction processes are labeled in Fig.~\ref{fig:Mandelstam}. $Q^2$ is the negative square of the four-momentum of the virtual photon, $Q^2 = -q^2$, and is quantifiable from the electron's final energy and scattering angle $\theta_{e'}$:
\begin{equation}
    Q^2 = 2E_eE_{e'}(1-\cos(\theta_{e'})),
\end{equation}
where $E_e$ and $E_{e'}$ are the energies of the initial and final-state electron, respectively. The backward DVCS cross section is expected to drop quickly with increasing $Q^2$. Therefore we often refer to this process as $u$-channel VCS rather than $u$-channel DVCS in order to not overstate the virtuality that may be expected of $u$-channel Compton scattering. The center-of-mass energy of the $p\gamma^*$ system is $W=\sqrt{s_{p\gamma^*}}=\sqrt{(p+q)^2}$, and is measurable through the momenta of the outgoing proton and photon. The Mandelstam $t=(p-p')^2$ is the square difference in four-momenta of the initial and final-state proton, and $u=(p-k)^2$ is the square difference in four-momenta of the beam proton and final-state photon. 

Two additional variables are often used to describe VCS in the $\gamma^*p$ center-of-mass frame. The scattering angle, $\theta$, is shown in Fig.~\ref{fig:COMFrame}. The momentum transfer involved in the scattering process is described equivalently in terms of $\theta$, $u$, or $t$. In this paper, we parameterize cross sections in terms of $u$ and $t$. Much of the literature quantifies measurements in terms of $\theta$~\cite{Laveissi_re_2009,PhysRevD.19.1921}, so it is often necessary to convert between the $\theta$ and Mandelstam parameterizations. $\phi$ is the azimuthal rotation of the final-state $\gamma p'$ plane with respect to the electron scattering plane. The $\phi$-dependence of the VCS cross section is related to orbital angular momentum contributions to proton GPDs~\cite{GPDs_article}.

We can construct the $ep\rightarrow e'p'\gamma$ cross section using these quantities:
\begin{equation}
    \frac{d^4\sigma[ep\rightarrow e'p'\gamma]}{dQ^2 dW d\phi dt} = \Gamma(Q^2,W)  \frac{d^2\sigma [\gamma^* p\rightarrow p'\gamma]}{d\phi dt}(Q^2,W,\phi,t),
\end{equation}
where $\Gamma(Q^2,W)$ is the virtual photon flux \cite{Budnev:1975poe}. Given a $\gamma^* p$ system with a defined $Q^2$ and $W$, the probability of the photon scattering with the proton to produce a final state with some $t$ and $\phi$ is thus proportional to $\frac{d^2\sigma}{d\phi dt}(Q^2,W,\phi,t)$. It is this reduced cross section for $\gamma^*p\rightarrow p'\gamma$ that is of primary interest to future EIC analyses. The form of this cross section is the subject of Sec.~\ref{section:themodel}.

It is often more convenient to discuss the cross section in terms of $u$ rather than the Mandelstam $t$, because the backward-production cross section as a function of $u$ behaves similarly to the forward-production cross section as a function of $t$. For this reason, we will also refer to the similar cross section:
\begin{equation}
    \frac{d^2\sigma[\gamma^*p\rightarrow p'\gamma]}{d\phi du}(Q^2,W,\phi,u).
\end{equation}

The Mandelstam $u$ is related to the scattering angle $\theta$ via:

\begin{equation}
\label{eq:costheta_u}
    \cos(\theta) = -\frac{G+2W^2(u-m_p^2)/(W^2-m_p^2)}{\sqrt{G^2-4W^2m_p^2}},
\end{equation}
where $G = m_p^2+Q^2+W^2$.
Equation~\ref{eq:costheta_u} can be rearranged to give $u$ in terms of the scattering angle:

\begin{equation}
\label{eq:u_costheta}
    u = m_p^2-\frac{W^2-m^2}{2W^2}\Big(G+\cos(\theta)\sqrt{G^2-4W^2m_p^2}\Big).
\end{equation}

Equation~\ref{eq:u_costheta} is used to compare our models with differential cross-section measurements at fixed scattering angles in Sec.~\ref{section:themodel}.
The most positive $u$ value is $u_0(Q^2,W) = u(Q^2,W,\cos\theta=-1)$, corresponding to $180^{\circ}$ backward Compton scattering.

A general relation from two-to-two particle scattering is useful here as well. For particles 1 and 2 scattering to produce particles 3 and 4, the Mandelstam relations give $s+t+u=m_1^2+m_2^2+m_3^2+m_4^2$. For VCS:
\begin{equation}
\label{tplusu}
    t+u = 2m_p^2 - Q^2 - W^2.
\end{equation}

This equation relates the $t$-dependence of the cross section to the $u$-dependence. At fixed $W$ and $Q^2$ the exponential rise of the cross section toward the most negative $t$ values in Fig.~\ref{fig:backwardspeak} can be translated into an exponential rise in the cross section toward the most positive possible $u$ values. Taken together, Equations~\ref{eq:u_costheta} and \ref{tplusu} relate the scattering angle to $t$. 

\section{Backward VCS Model}

As discussed in Sec.~\ref{section:generalkinematics}, VCS may be described by the kinematic variables: $Q^2$, $W$, $u$ (or $t$), and $\phi$. 
The event-plane rotation in azimuth does not affect the feasibility of detecting VCS events, so for this paper, all cross sections and rates are integrated over $\phi$. 

DVCS off of protons is often modeled with a differential cross section of the form:
\begin{equation}
\label{eq:dsigdt}
    \frac{d\sigma}{dt}\sim\exp(-B|t|).
\end{equation}
Exclusive vector-meson production can be modeled by the same functional form, in agreement with data at forward angles.
This dependence is a somewhat simplified picture as was demonstrated by the H1 Collaboration, which measured the $B$ parameter and showed it to vary slowly with $Q^2$~\cite{H1:1999pji}. 
The DVCS event-generator MILOU allows users to express $B$ as a linear function of $\ln Q^2$~\cite{Perez:2004ig}, and eSTARlight expresses cross sections for vector-meson production on protons in terms of a dipole form factor~\cite{PhysRevC.99.015203} that reduces to Eq.~\ref{eq:dsigdt} as $t\rightarrow0$. 

Despite their differences, these parameterizations all include a sharp peak at $t\sim0$ and a vanishing cross section as $|t|$ rises. 
However, early photoproduction data found a breakdown of these parameterizations at very large $|t|$~\cite{Clifft:1977yi, PhysRevLett.23.725}. 
Instead of an ever-decreasing cross section with increasing $|t|$, an enhancement was observed at the maximal $|t|$ values. This is interpreted as coming from contributions from baryon (Reggeon) exchange trajectories.
Recent measurements, seen in Fig.~\ref{fig:backwardspeak}, extend $u$-channel $\omega$ electroproduction data to high $Q^2$~\cite{billspaper}. 
However, a peak in the VCS cross section at maximal $|t|$ has not yet been observed, likely due to the challenges of detecting backward VCS in fixed-target experiments.  
Experiments have been proposed at Jefferson Lab to establish the existence of this peak \cite{Li:2022dxk}. 

We exploit expected similarities between the $t$ and $u$-channel exchanges and model the backward cross section with the form:
\begin{equation}
\label{eq:exp}
  \frac{d\sigma}{du}(u) \sim \exp(-D|u-u_{\text{0}}|),
\end{equation}
which describes an exponentially falling cross section as $u$ deviates from its most positive possible value $u_0$. There exists little data on VCS at backward angles to constrain the value of $D$, often called the ``slope parameter." We can estimate $D$ using data from backward production of $\omega$ mesons. 

Daresbury Laboratory's NINA 5~GeV electron synchrotron~\cite{Clifft:1977yi} and Jefferson Lab Hall~C~\cite{billspaper} have both measured the $u$-channel peak in backward $\omega$ production. The NINA measurements correspond to photoproduction ($Q^2$=0) and the Hall C measurements are for electroproduction at $Q^2=1.75$~GeV$^2$ and 2.45~GeV$^2$. The Hall~C data had fewer bins in $u$, measuring the cross section at three $u$ values for each set of kinematics, compared to around twenty measurements in $u$ from NINA for each configuration. The Hall~C and NINA data are at similar values of $u$ and $W$, and differ primarily in their $Q^2$ and the number of measurements in $u$. The NINA measurements show a steep exponential drop-off near $|u|\sim0$ and a slower drop-off at large $|u|$, with a dip in between. The Hall~C measurement does not show a dip in the cross section, and is well-described by the simple exponential of Eq.~\ref{eq:exp}. The slope parameter obtained from the Hall~C data is $D=2.4\pm1.8$~GeV$^{-2}$~\cite{billspaper} at $W=2.5$~GeV and $0.03<-u<0.28$~GeV$^2$. The slope parameter does not depend on $Q^2$ over the measured range. The steep $u$-exponential portion of the NINA photoproduction cross section was fit for 4.7~GeV photons ($W=3.1$~GeV), resulting in a slope parameter of $D=21.8\pm1.2$~GeV$^{-2}$, valid over $0.01<-u<0.12$~GeV$^2$. 

It is not clear what causes the difference between the Hall~C and NINA slope parameters. Photoproduction of $\omega$ mesons might have a very different behavior than electroproduction at low $|u|$. Another possibility is that a steep slope and dip structure are integrated over in the Hall~C data or that these features are not present at the lower $W$ measured by Hall~C. Additional electroproduction measurements with fine binning in $u$ are needed for a decisive comparison. In the absence of backward VCS measurements, we use these very different slopes to develop two alternate models of backward VCS (referred to as models 1 and 2) that, taken together, provide us with a sense for the range of possible cross sections and kinematics. 

Model~1 uses the $D=2.4$~GeV$^{-2}$ value measured by Hall~C. Model~2 uses the $D=21.8$~GeV$^{-2}$ value from the NINA data. In a vector-meson-dominance framework, it is not unreasonable to assume that $u$-channel VCS would behave similarly to $u$-channel $\omega$ production. We use these two slope-parameter values for our backward VCS models in lieu of more representative data. In order to minimize uncertainties on the cross section caused by these two different slope parameters, we describe below how both cross section models are scaled to the limited backward VCS data that is available.

We next model the $W$-dependence of the cross section. Backward-production cross sections scale with a negative power of the center-of-mass energy $W$, representing the scaling behavior of Reggeon exchange trajectories~\cite{Crittenden:1997yz,Klein:1999qj}:
\begin{equation}
    \sigma_{\gamma^* p\rightarrow Xp}(W) \sim W^{-\eta}.
    \label{eq:sigma}
\end{equation} 

There is currently no data on the $W$-dependence of the backward VCS cross section at both fixed $u-u_0$ and fixed $Q^2$. A reasonable starting point is $d\sigma/du(W)\sim (W^2-m_p^2)^{-2}$~\cite{Laveissi_re_2009,PhysRevD.79.033012}, which has also been used to model backward meson production~\cite{billspaper}. This is similar to the $(W^2-m_p^2)^{-2.7}$ dependence previously used in backward vector-meson simulations~\cite{PhysRevC.106.015204}. 

We employ a squared nucleon dipole form factor for the explicit $Q^2$-dependence. This goes as $\sim(Q^2 + \Lambda^2)^{-4}$ for some constant $\Lambda$.
Backward VCS data were collected at a constant scattering angle at $W$=1.53~GeV in the resonance region, where hadronic resonances give structure to the cross section in $W$. A fit to this data using the dipole form-factor scaling found $\Lambda^2=2.77\ \text{GeV}^2$~\cite{Laveissi_re_2009}. This is the best measurement of the $Q^2$ scaling of backward VCS, but it should be noted that above the resonance region the cross section may scale differently. For example, in backward $\omega$ production at $W=2.21$~GeV, the cross section goes as $\sim$$Q^{-1.08}$ for transversely-polarized photons and as $\sim$$Q^{-10.22}$ for longitudinally-polarized photons~\cite{PhysRevLett.123.182501}. Our models use the $(Q^2+2.77\ \text{GeV}^2)^{-4}$ scaling, which is representative of the data that is most relevant here. Combining this scaling with the $W$ scaling, the backward VCS cross-section model is:

\begin{equation}
\label{genericXsec}
\frac{d\sigma}{du}(Q^2,W,u) \approx \frac{A\exp(-D|u-u_{\text{0}}|)}{(W^2-m_p^2)^\alpha(Q^2+\Lambda^2)^4/\text{GeV}^{12}},
\end{equation}
where $\alpha=2$, $\Lambda^2 = 2.77$~GeV$^2$, and $A$ is a normalization factor. The Jefferson Lab Hall~A Collaboration's VCS data~\cite{Laveissi_re_2009} at fixed angle ($\cos\theta\ =\ -0.975$), and $Q^2$\ =\ 1~GeV$^2$ are used to anchor the cross-section amplitude. 
To limit effects of nucleon resonances that decay to $\gamma p$, the model amplitudes were fit to cross section measurements at the eleven highest $W$ values, from 1.77 to 1.97~GeV. 
Statistical and systematic uncertainties on the data points were combined in quadrature. 

\label{section:themodel}
\begin{figure}
  \begin{center}
    \includegraphics[width=0.5\textwidth]{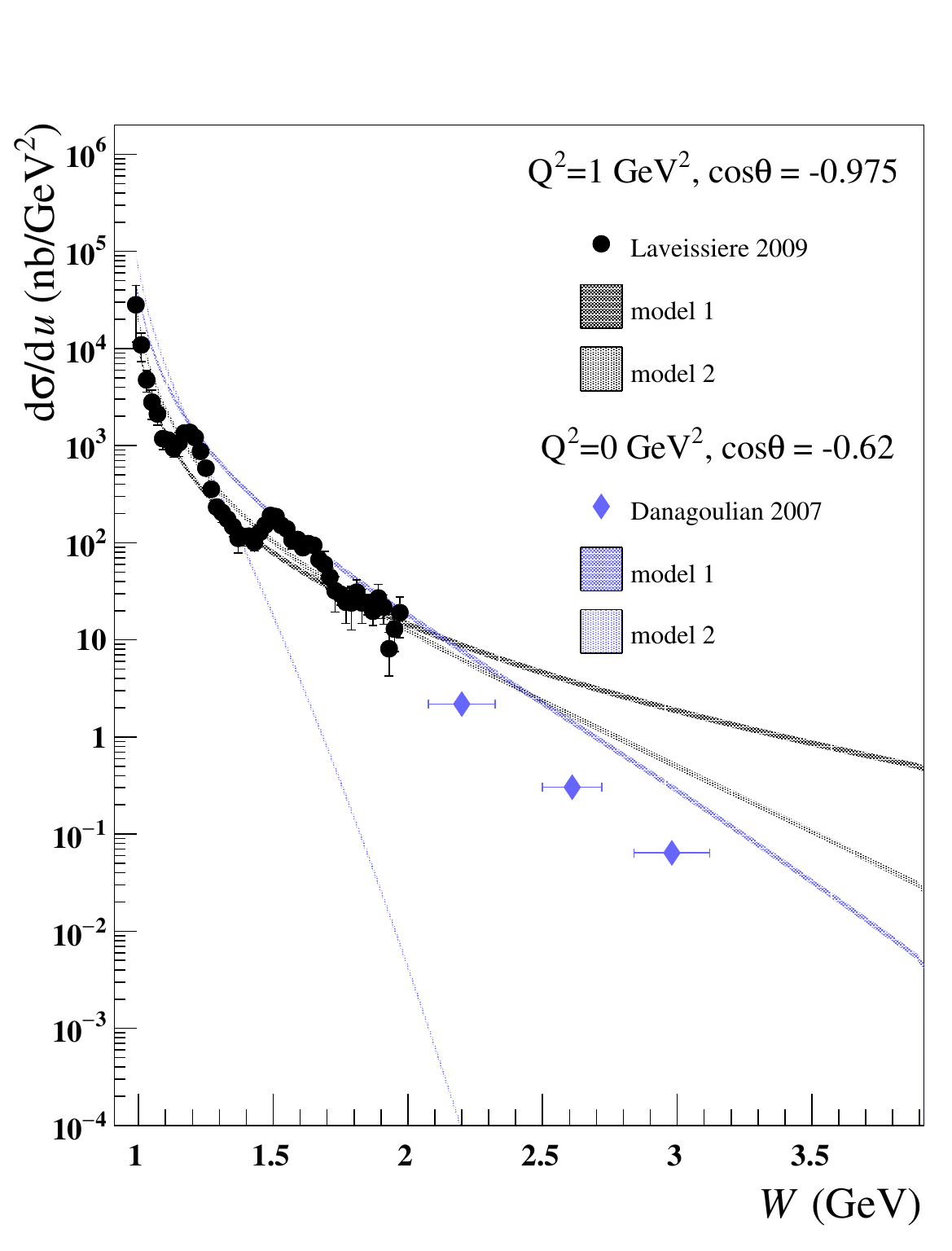}
    \caption{Differential cross sections for backward virtual and real Compton scattering as a function of $W$. Models~1 and 2 are each compared with backward VCS data at $Q^2=1$~GeV$^2$ and $\cos\theta=-0.975$~\cite{Laveissi_re_2009}, and with near-backward RCS data at $Q^2=0$~GeV$^2$ and $\cos\theta = -0.62$~\cite{PhysRevLett.98.152001_Dan}.}
    \label{fig:xseccomparison}
  \end{center}
\end{figure}

The fit finds an amplitude of $A=32$~$\mu$b/GeV$^{2}$ for model~1 and $A=65$~$\mu$b/GeV$^{2}$ for model~2, each with a 7$\%$ uncertainty from the normalization fit. 
The two models are compared in Fig.~\ref{fig:xseccomparison} along with the VCS data. 
These measurements were taken at constant scattering angles, $\theta$. 
For the comparison to data, both $u$ and $u_0$ were calculated at each $W$ value, so the differential cross sections shown in Fig.~\ref{fig:xseccomparison} do not have the simple $\sim(W^2-m_p^2)^{-2}$ dependence as might be expected. The parameters for both models and the background model discussed in Sec.~\ref{section:backgrounds} are summarized in Tab.~\ref{tab:models}. The table also summarizes the kinematic ranges over which the data informing each parameter were taken. 

With the amplitudes fixed, each model was then used to predict the differential cross section as a function of $W$ for wide-angle real Compton scattering (RCS), and compared with data~\cite{PhysRevLett.98.152001_Dan} at $\cos\theta=-0.62$.
Models 1 and 2 both demonstrate plausible scaling behavior when compared to the backward-angle VCS data. 
For the RCS data at the wide backward angle, model~1 performs significantly better, although it still overshoots the data. 
Although the RCS data with $\cos\theta=-0.62$ corresponds to backward scattering, it is not close to the backward peak, and may not be well-described by the simple exponential $u$-dependence employed at the most backward angles.
It is not surprising then that neither model matches the data, but the near-agreement of model~1 is a good reason to move forward with this model.
We therefore use model~1 ($D=2.4$~GeV$^{-2}$, $A=32$~$\mu$b/GeV$^{2}$) in the simulations described in Sec.~\ref{section:simulations}. 

\begin{table*}
\begin{tabular}{| c c c c c c c c c c c c c |}
\hline
{}&{}&{}&{    }&{}&{     }&\multicolumn{7}{c|}{Informed by}\\
\cline{7-13}
{model}&{  }&{parameter}&{  }&{value}&{ }&{production of}&{\hspace*{15pt}}&{$Q^2$ (GeV$^2$)}&{\hspace*{10pt}}&{$W$ (GeV)}&{\hspace*{10pt}}&{$-u$ (GeV$^2$)}\\
\hline 
\multirow{ 4}{*}{VCS model 1}&{  }&{$A$}&{}&{32~$\mu$b/GeV$^2$}&\hspace*{20pt}&{$\gamma$}&{ }&{1}&{ }&{[2.8,3.0]}&{ }&{0.1}\\
&{ }&{$D$}&{ }&{2.4~GeV$^{-2}$}&{ }&{$\omega$}&{ }&{[1.8,2.5]}&{ }&{2.5}&{ }&{[0.0,0.3]}\\
&{  }&{$\alpha$}&{ }&{2}&{ }&{$\gamma$}&{ }&\multicolumn{5}{c|}{--------- theoretical scaling ---------}\\
&{  }&{$\Lambda^2$}&{ }&{2.77~GeV$^2$}&{ }&{$\gamma$}&{ }&{[0.6,2.0]}&{ }&{1.5}&{ }&{[0.5,0.6]}\\
\hline 
\multirow{ 4}{*}{VCS model 2}&{  }&{$A$}&{}&{65~$\mu$b/GeV$^2$}&\hspace*{20pt}&{$\gamma$}&{ }&{1}&{ }&{[2.8,3.0]}&{ }&{0.1}\\
&{ }&{$D$}&{ }&{21.8~GeV$^{-2}$}&{ }&{$\omega$}&{ }&{0}&{ }&{3.1}&{ }&{[0.0,0.1]}\\
&{  }&{$\alpha$}&{ }&{2}&{ }&{$\gamma$}&{ }&\multicolumn{5}{c|}{--------- theoretical scaling ---------}\\
&{  }&{$\Lambda^2$}&{ }&{2.77~GeV$^2$}&{ }&{$\gamma$}&{ }&{[0.6,2.0]}&{ }&{1.5}&{ }&{[0.5,0.6]}\\
\hline 
\multirow{ 4}{*}{$\pi^0$ model}&{  }&{$A$}&{}&{1.26~mb/GeV$^2$}&\hspace*{20pt}&{$\pi^0$}&{ }&{0}&{ }&{3.5}&{ }&{[0.0,0.3]}\\
&{ }&{$D$}&{ }&{4.2~GeV$^{-2}$}&{ }&{$\pi^0$}&{ }&{0}&{ }&{[3.5,5.9]}&{ }&{[0.0,0.3]}\\
&{  }&{$\alpha$}&{ }&{2.8}&{ }&{$\pi^0$}&{ }&{0}&{ }&{[3.5,5.9]}&{ }&{[0.0,0.5]}\\
&{  }&{$\Lambda^2$}&{ }&{2.77~GeV$^2$}&{ }&{$\pi^0$}&{ }&{[0.9,1.2]}&{ }&{2.0}&{ }&{[0.0,0.4]}\\
\hline
\end{tabular}
\caption{Summary of $u$-channel VCS and $\pi^0$ model parameters using the generic model given by Eq.~\ref{genericXsec}. The rightmost four columns summarize the kinematics of the data that were used to motivate these parameter values. For example, the slope parameter $D=2.4$~GeV$^{-2}$ used in VCS model~1 is motivated by $\omega$ production data at $1.8<Q^2<2.5$~GeV$^2$, $W=2.5$~GeV, and $0.0<-u<0.3$~GeV$^2$.}
\label{tab:models}
\end{table*}

\section{Backgrounds}
\label{section:backgrounds}

\begin{figure}
  \begin{center}
    \includegraphics[width=0.4\textwidth]{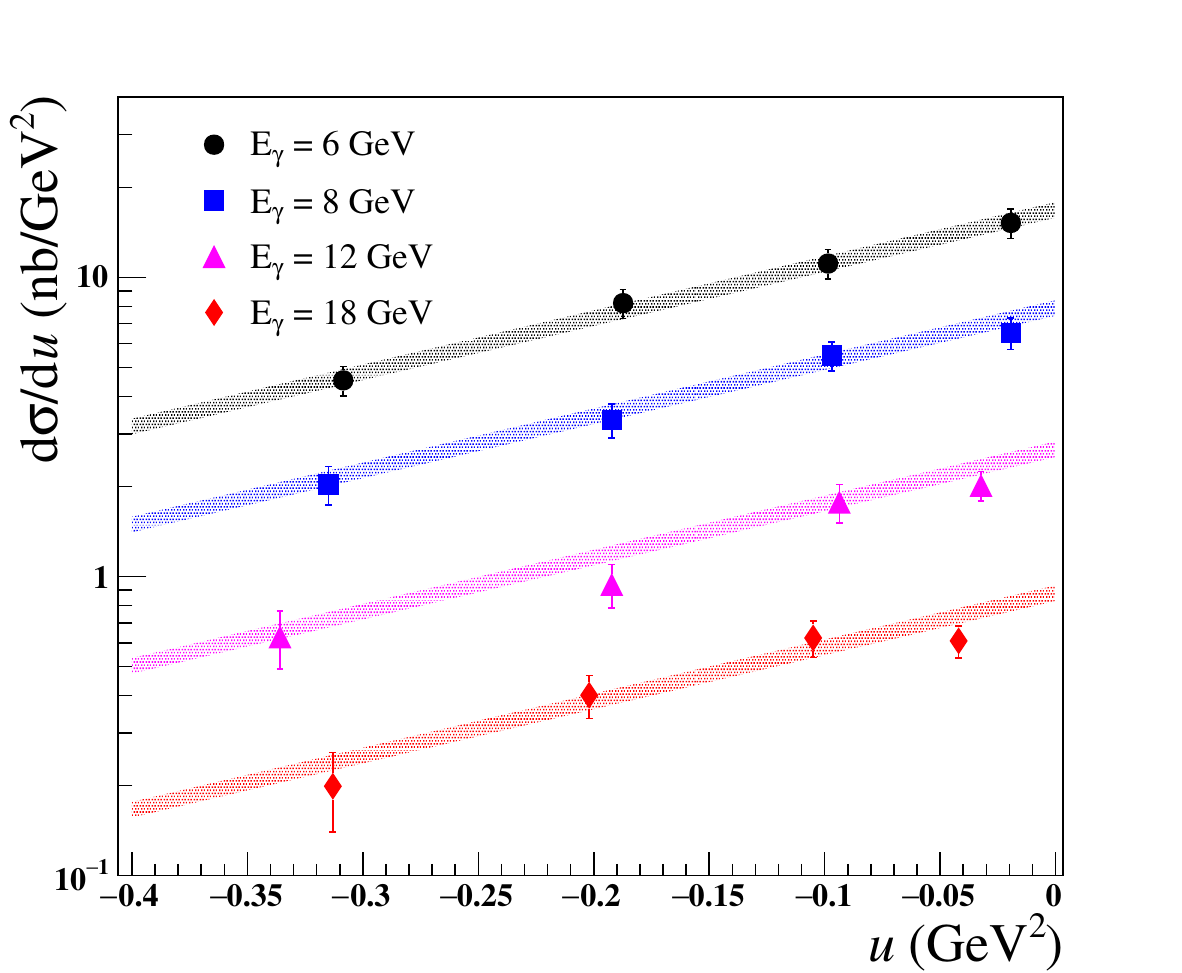}
    \caption{$\gamma p\rightarrow\pi^0 p'$ model cross section (shaded bands) comparison with photoproduction data from Ref.~\cite{PhysRevLett.23.725}.}
    \label{fig:pi0xsecs}
  \end{center}
\end{figure}
\begin{figure}
  \begin{center}
    \includegraphics[width=0.4\textwidth]{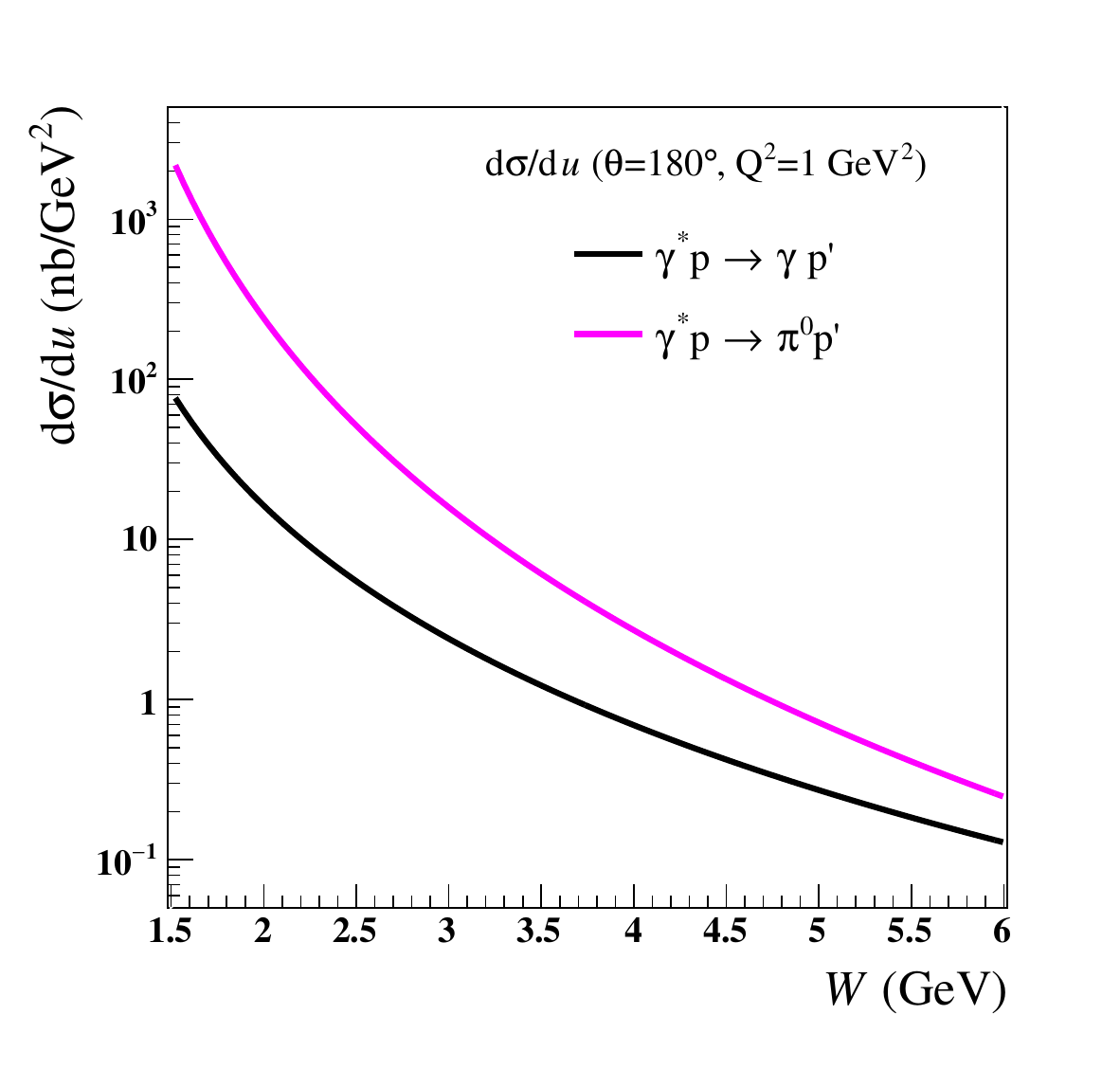}
    \caption{Model comparison of $\pi^0$ production and Compton scattering cross sections at $Q^2$=1~GeV$^2$.}
    \label{fig:pi0dvcscomparison}
  \end{center}
\end{figure}

Backgrounds in VCS measurements are dominated by the Bethe-Heitler process and $\pi^0$ production.
However the Bethe-Heitler cross section, which peaks in the forward region, does not have a complementary backward peak. Thus there is almost no Bethe-Heitler contribution to the background for backward Compton scattering~\cite{Pire:2021hbl, LANSBERG200716}.
The $\pi^0\rightarrow\gamma\gamma$ process does have a peak in the backward limit~\cite{PhysRevLett.23.725} and is expected to be the primary background to $u$-channel VCS.

In Sec.~\ref{section:simulatedbackground} we discuss the separation of backward $\pi^0$ production from backward Compton events. The model used to generate these $u$-channel $\pi^0$ events is similar to the backward VCS model. Unlike Compton scattering, a backward peak in the cross section has already been observed for $\pi^0$ production in fixed-target experiments at the Stanford Linear Accelerator~\cite{PhysRevLett.23.725}. 
At photon energies between 6 and 18~GeV, the cross section was found to scale as $d\sigma/du\sim W^{-6.0\pm0.4}$. We performed a fit to the data assuming a scaling of the form $d\sigma/du\sim (W^2-m_p^2)^{-\alpha}$, which described the data well with $\alpha = 2.8\pm 0.1$.

Similar to backward VCS, the backward $\pi^0$ production cross section may be expected to scale with $Q^2$ according to a squared nucleon dipole form factor: $\sim(Q^2+\Lambda^2)^{-4}$. Jefferson Lab's Hall~A Collaboration measured this $Q^2$-dependence of backward $\pi^0$ production~\cite{PhysRevC.69.045203} which, at $W=2$~GeV, was found to be consistent with the $\Lambda^2=2.77~\rm{GeV}^2$ measured in backward VCS~\cite{Laveissi_re_2009}. We therefore use the same $Q^2$ dependence in our model of the $u$-channel $\pi^0$ cross section. Taken together, these scalings lead to the following model for backward $\pi^0$ production:
\begin{equation}
\frac{d\sigma}{du}(Q^2,W,u) \approx \frac{A\exp(-D|u-u_{\text{0}}|)}{(W^2-m_p^2)^{2.8}(Q^2+\Lambda^2)^{4}/\text{GeV}^{12}}.
\end{equation}

In order to extract the $A$ and $D$ parameters, we performed fits to the SLAC backward $\pi^0$ production data~\cite{PhysRevLett.23.725}, accounting for the $W$ and $Q^2$ scalings. The fits were done in the region $-0.34$$< $$u $$ <$0.0~GeV$^2$, over which the cross section is nicely described by the exponential behavior in $u$.  The amplitude $A$ and slope $D$ were found to be $1.26 \pm 0.07$~mb/GeV$^2$ and $4.2\pm 0.4$~GeV$^{-2}$ respectively. A comparison of the model to the photoproduction data is shown in Fig.~\ref{fig:pi0xsecs}.

\section{Simulating Backward Compton Scattering and $\pi^0$ Production}
\label{section:simulations}

Backward VCS and $\pi^0$-production simulations were performed using the eSTARlight Monte Carlo event generator
\cite{PhysRevC.99.015203}, which was modified by the authors to include these processes with the kinematics described in Sections~\ref{section:themodel} and \ref{section:backgrounds}. eSTARlight was developed for modeling exclusive meson production in $ep$ and $eA$ collisions, and thus already includes much of the framework needed for simulating backward Compton scattering and $\pi^0$ production.

\begin{figure}
  \begin{center}
    \includegraphics[width=0.49\textwidth]{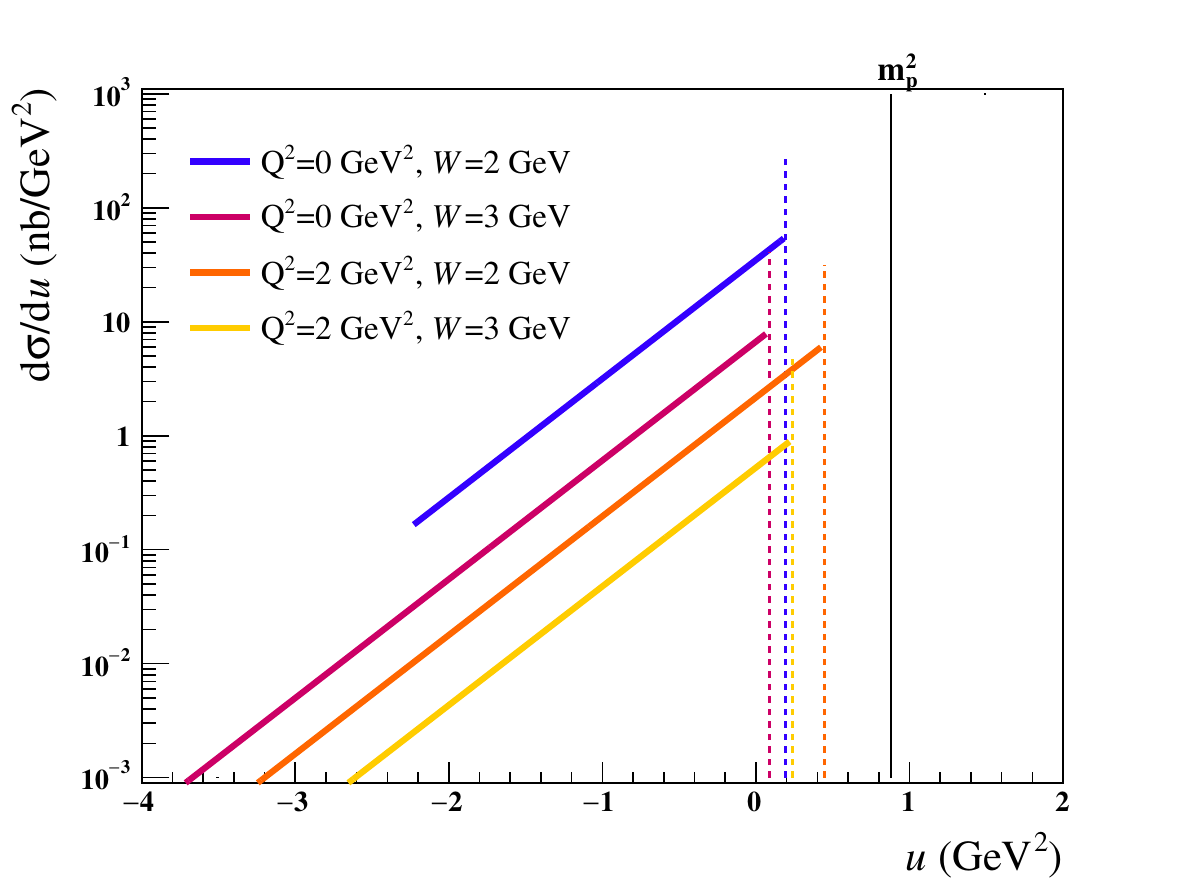}
    \caption{Comparison of differential cross sections at fixed $Q^2$ and $W$ demonstrating exponential dependence, with maxima at varying $u_0$ (dotted vertical lines).}
    \label{fig:multiplexsecs}
  \end{center}
\end{figure}

The code first generates a virtual photon spectrum using a lookup table, representing the photon flux $\Gamma(k,Q^2) = d^2N_{\gamma^*}/dkdQ^2$~\cite{PhysRevC.99.015203}. Here $k$ refers to the energy of the virtual photon. For each $k$ and $Q^2$ selected, the center-of-mass energy, $W$, is calculated. The generated photon is then compared against the cross section models for backward VCS or backward $\pi^0$ production, $d\sigma/du(Q^2,W)$, and is accepted or rejected by Monte Carlo sampling. After the virtual photon kinematics have been selected, $u_0$ is calculated and the Mandelstam $u$ of the process is then selected according to random sampling from the $d\sigma/du(u)\sim\exp{(-D|u-u_{\text{0}}|)}$ distribution. 
Figure~\ref{fig:multiplexsecs} shows several example exponential differential cross sections using model~1 for a given $W$ and $Q^2$. The $u_0$ for each set of kinematics, as calculated using Eq.~\ref{eq:costheta_u}, is shown as a vertical dashed line corresponding to each cross section.

The event is treated in the center-of-mass frame as in Fig.~\ref{fig:COMFrame}. 
For a given value of $u$, the polar angle $\theta$ of the outgoing photon with respect to the initial-state axis is given by Eq.~\ref{eq:costheta_u}.
With $\theta$ chosen, the next step is to generate a value of the azimuthal rotation $\phi$, which is uniformly sampled.
For $\pi^0$ production, the $\pi^0$ is then decayed isotropically via $\pi^0\rightarrow\gamma\gamma$.
Finally the event is rotated and boosted back to the laboratory frame, and the final-state particles are written to file.

\begin{table*}
\begin{tabular}{| c c c c c c c c c c c c c c c |}
\hline
{}&{}&{}&{    }&\multicolumn{3}{c}{Model 1}&{     }&\multicolumn{3}{c}{Model 2}&{     }&\multicolumn{3}{c|}{$\pi^0$ Production} \\
\cline{5-7}
\cline{9-11}
\cline{13-15}
{Collision}&{  }&{$Q^2$ range}&{  }&{$\sigma_\text{tot}$}&{ }&{events}&{     }&{$\sigma_\text{tot}$ }&{ }&{events}&{     }&{$\sigma_\text{tot}$ }&{ }&{events}\\
{energy}&{  }&{(GeV$^2$)}&{  }&{(pb)}&{ }&{per 10 fb$^{-1}$}&{     }&{(pb)}&{ }&{per 10 fb$^{-1}$}&{     }&{(pb)}&{ }&{per 10 fb$^{-1}$}\\
\hline 
\multirow{ 4}{*}{5$\times$41 GeV} &{  }     & 0 - 5 &\hspace*{30pt}& 11  & & 1.1$\times$10$^{5}$  &\hspace*{30pt}& 2.4 &  & 2.4$\times$10$^{4}$   &\hspace*{30pt}& 69 &  & 6.9$\times$10$^{5}$ \\
&{  }& $10^{-3}$ - 1 &{  }& 2.7  & & 2.7$\times$10$^{4}$  && 0.60 &  & 6.0$\times$10$^{3}$   && 17 &  & 1.7$\times$10$^{5}$ \\
&{  }& 1 - 2 &{  }  & 2.2$\times$10$^{-2}$  & & 220  && 4.7$\times$10$^{-3}$ &  & 47 && 8.8$\times$10$^{-2}$  &  & 880 \\
&{  }& 2 - 5 &{  } & 1.9$\times$10$^{-3}$  & & 19  && 4.1$\times$10$^{-4}$ &  & 4.1  && 4.5$\times$10$^{-3}$  &  & 45 \\
\hline 
\multirow{ 4}{*}{10$\times$100 GeV} &{  } & 0 - 5 &{  }& 12  & & 1.2$\times$10$^{5}$  && 2.7 &  & 2.7$\times$10$^{4}$   && 79 &  & 7.9$\times$10$^{5}$ \\
&{  }& $10^{-3}$ - 1 &{  }& 2.8 & & 2.8$\times$10$^{4}$  && 0.61 &  & 6.1$\times$10$^{3}$   && 17 &  & 1.7$\times$10$^{5}$ \\
&{  }& 1 - 2  &{  }& 2.2$\times$10$^{-2}$  & & 220 && 4.8$\times$10$^{-3}$ &  & 48   && 8.9$\times$10$^{-2}$  &  & 890 \\
&{  }& 2 - 5 &{  }& 1.9$\times$10$^{-3}$  & & 19 && 4.2$\times$10$^{-4}$ &  & 4.2  && 4.6$\times$10$^{-3}$  &  & 46 \\
\hline 
\multirow{ 4}{*}{18$\times$275 GeV} &{  }& 0 - 5 &{  }& 14  & & 1.4$\times$10$^{5}$  && 3.1 &  & 3.1$\times$10$^{4}$   && 89 &  & 8.9$\times$10$^{5}$ \\
&{  }& $10^{-3}$ - 1 &{  }& 2.8 & & 2.8$\times$10$^{4}$  && 0.61 &  & 6.1$\times$10$^{3}$   && 17 &  & 1.7$\times$10$^{5}$ \\
&{  }& 1 - 2 &{  }& 2.2$\times$10$^{-2}$  & & 220 && 4.8$\times$10$^{-3}$ &  & 48   && 8.9$\times$10$^{-2}$  &  & 890 \\
&{  }& 2 - 5 &{  }& 1.9$\times$10$^{-3}$  & & 19 && 4.2$\times$10$^{-4}$ &  & 4.2  && 4.6$\times$10$^{-3}$  &  & 46 \\
\hline
\end{tabular}
\caption{Cross sections and total number of events per 10 fb$^{-1}$ of integrated luminosity with $W>2$~GeV for $u$-channel VCS in models~1 and 2 and $u$-channel $\pi^0$ production.}
\label{tab:rates}
\end{table*}

The models considered in Sec.~\ref{section:themodel} include a divergence as $W\rightarrow m_p$, which is consistent with expectations. Compton scattering does not have a lower threshold on the energy of produced photons, so there is no restriction on $W$ approaching the proton mass. However, it becomes increasingly difficult to discriminate forward and backward scattering for Compton scattering at low $W$. From Eq.~\ref{eq:u_costheta}, the difference between $u_{\text{f}}$ for forward scattering at $\cos\theta=1$ and $u_{\text{b}}$ representing backward scattering at $\cos\theta=-1$ vanishes at threshold. We therefore place a lower-limit on $W$ at 2~GeV, which also limits contributions from the resonance region.

We consider three standard EIC collision energies: 5~GeV electrons on 41~GeV protons (5$\times$41~GeV), 10~GeV electrons on 100~GeV protons (10$\times$100~GeV), and 18~GeV electrons on 275~GeV protons (18$\times$275~GeV). 
These three energy combinations will allow us to study how predicted acceptances for the planned ePIC detector will compare to the phase space of final-state particles from backward VCS and $\pi^0$ production.

\section{Detection of Backward VCS}
\label{section:detection}
\subsection{Relative Rates and Kinematics for Backward VCS}
\label{section:rates}

Models~1 and 2 were scaled according to VCS data~\cite{Laveissi_re_2009} as described in Sec.~\ref{section:themodel}. The backward $\pi^0$ production model was likewise scaled to fit existing data in Sec.~\ref{section:backgrounds}. When combined with the virtual photon flux, these models can be used to calculate the total expected cross section (integrated over $W$, $Q^2$, and $u$) for a given collider configuration. These total cross sections are given in Tab.~\ref{tab:rates} along with the predicted number of events per 10~fb$^{-1}$ of integrated luminosity.

The event rates do not increase significantly at higher collision energies.
This is because the $u$-channel cross sections are modeled via Regge exchange, which - unlike Pomeron exchange - are characterized by a rapid decrease with increasing $W$. The virtual photon spectrum scales as roughly $1/k$, so the addition of a small high-energy cross section does not significantly increase the total event rate. 

The steep decrease in rates with increasing $Q^2$ suggests that these processes are only measurable at low $Q^2$.  
However, this conclusion is highly model dependent, and the cross-section dependence on the photon polarization may have the effect of moderating its decline at large $Q^2$, as mentioned in Sec.~\ref{section:themodel}. A more moderate scaling may allow detectable transversely-polarized backward VCS interactions at moderate $Q^2$ values, so the rates in Tab.~\ref{tab:rates} should not discourage studies of $u$-channel VCS at moderate $Q^2$.

\subsection{$u$-Channel VCS Simulations}
\label{section:simulatedvcs}

\begin{figure*}
  \begin{center}
    \includegraphics[width=0.9\textwidth]{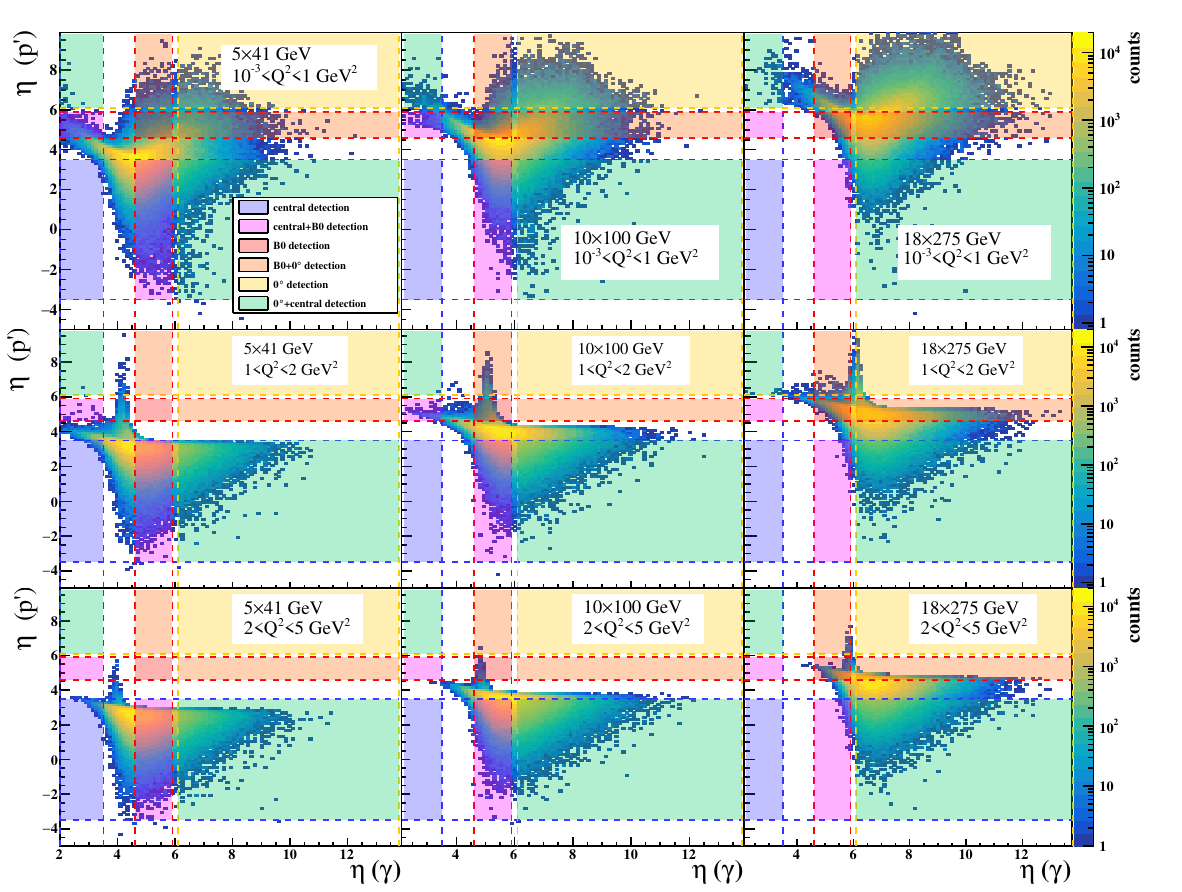}
\caption{Pseudorapidity distributions of final-state proton-photon pairs for $u$-channel Compton scattering using model~1 with $W>2$~GeV. The shaded colored regions show the acceptance in pseudorapidity of the central, B0, and zero-degree detectors.}
    \label{fig:energyscan}
  \end{center}
\end{figure*}

\begin{table}[b]
\begin{tabular}{| c c c c c c c c |}
\hline
\multicolumn{1}{|c}{Proton}&{B0}&{  }&\multicolumn{5}{c|}{$p'\gamma$ geometric acceptance}\\
\multicolumn{1}{|c}{beam}&{EMCal.}&{  }&{$10^{-3}$$<$$Q^2$$<$$1$}&{  }&{$1$$<$$Q^2$$<$$2$}&{  }&{$2$$<$$Q^2$$<$$5$}\\
\hline 
\multirow{ 2}{*}{41 GeV}& no&  & 2.9\% &  & 5.6\% &  & 6.1\% \\
&yes&  & 27\% &  & 62\% &  & 66\% \\
\hline 
\multirow{ 2}{*}{100 GeV}& no&  & 17\% &  & 8.2\% &  & 23\% \\
&yes&  & 41\% &  & 20\% &  & 59\% \\
\hline 
\multirow{ 2}{*}{275 GeV}& no&  & 75\% &  & 65\% &  & 27\% \\
&yes&  & 81\% &  & 71\% &  & 29\% \\
\hline
\end{tabular}
\caption{Acceptances for $p'+\gamma$ detection for the ePIC/ECCE design~\cite{Adkins:2022jfp} and VCS model~1 with $W>2$~GeV. $Q^2$ ranges are in units of GeV$^2$.}
\label{tab:geoeff}
\end{table}

The initial EIC detector, the Electron-Proton/Ion Collider Experiment (ePIC), evolved from it's predecessor, the ECCE proposal~\cite{Adkins:2022jfp}. Optimization of the ePIC design is ongoing, so we use here the detector described in the ECCE proposal, which should be close to ePIC. The ATHENA detector proposal~\cite{Adam_2022} had similar acceptances and would have had a similar performance for backward Compton scattering.

ePIC will include a central region consisting of charged-particle tracking and calorimetry covering an acceptance of $-3.5<\eta<3.5$.
An additional subsystem will be embedded within a dipole magnet referred to as the B0 steering dipole located about 6~m downstream from the central detector region.  The B0 magnet encloses the ion and electron beam pipes and includes a cavity in which charged-particle trackers and an electromagnetic calorimeter may be embedded, extending the acceptance to roughly $4.6<\eta<5.9$.
In the far-forward regime, a zero-degree calorimeter (ZDC) provides photon-detection capabilities for approximately $\eta>6.1$. 
Roman pots covering nearly the same range will be able to detect minimally-scattered protons~\cite{Adkins:2022jfp}.
Both the ZDC and a B0 calorimeter will be critical for detecting the far-forward photons produced in backward Compton scattering and $\pi^0$ production.
Due to the non-zero beam crossing angle and constraints on the detector designs, the B0, ZDC, and Roman pots will not be not azimuthally symmetric in ePIC, so the acceptances taken here are only approximations.  

In Sec.~\ref{section:simulatedbackground}, the use of exclusivity cuts to reduce the backward $\pi^0$ background is explored. These cuts require reconstruction of the entire event, including the scattered electron. The EIC low-$Q^2$ taggers will have acceptance for electrons with $Q^2$ down to $10^{-7}$~GeV$^2$~\cite{ABDULKHALEK2022122447}. However the angular divergence of the electron beam makes reconstruction imprecise for $Q^2<10^{-3}$~GeV$^2$ and impossible for $Q^2<10^{-4}$~GeV$^2$. An additional complication is expected from minimally-scattered bremsstrahlung electrons that will be difficult to separate from low-$Q^2$ electrons. The effect of these electrons on low-$Q^2$ reconstruction will depend on the luminosity and design specifics of the electron taggers as discussed, for example, in the ATHENA proposal~\cite{Adam_2022}.  We therefore only simulate events with $Q^2$ greater than $10^{-3}$~GeV$^2$, although the majority of the cross section is below this threshold.

Figure~\ref{fig:energyscan} shows acceptance plots for the final-state proton and photon from simulations of $10^6$ backward Compton scattering events in eSTARlight. These are for $ep$ collisions at 5$\times$41~GeV, 10$\times$100~GeV, and 18$\times$275~GeV. The acceptances are shown in three kinematic regimes: nearly-real ($10^{-3}<Q^2<1~\text{GeV}^2$), virtual ($1<Q^2<2~\text{GeV}^2$), and deeply-virtual Compton scattering ($2<Q^2<5~\text{GeV}^2$). Superimposed on the scatter plots are the approximate pseudorapidity coverages of the ePIC central detectors, B0 detectors, and the zero-degree detectors (ZDC and Roman pots).

In backward VCS events at high $Q^2$, the proton is often shifted far enough in rapidity to be detectable by the central charged-particle trackers. At low $Q^2$, backward Compton scattering measurements will rely on B0 tracking to observe the proton. 

\begin{figure*}
  \begin{center}
    \includegraphics[width=0.8\textwidth]{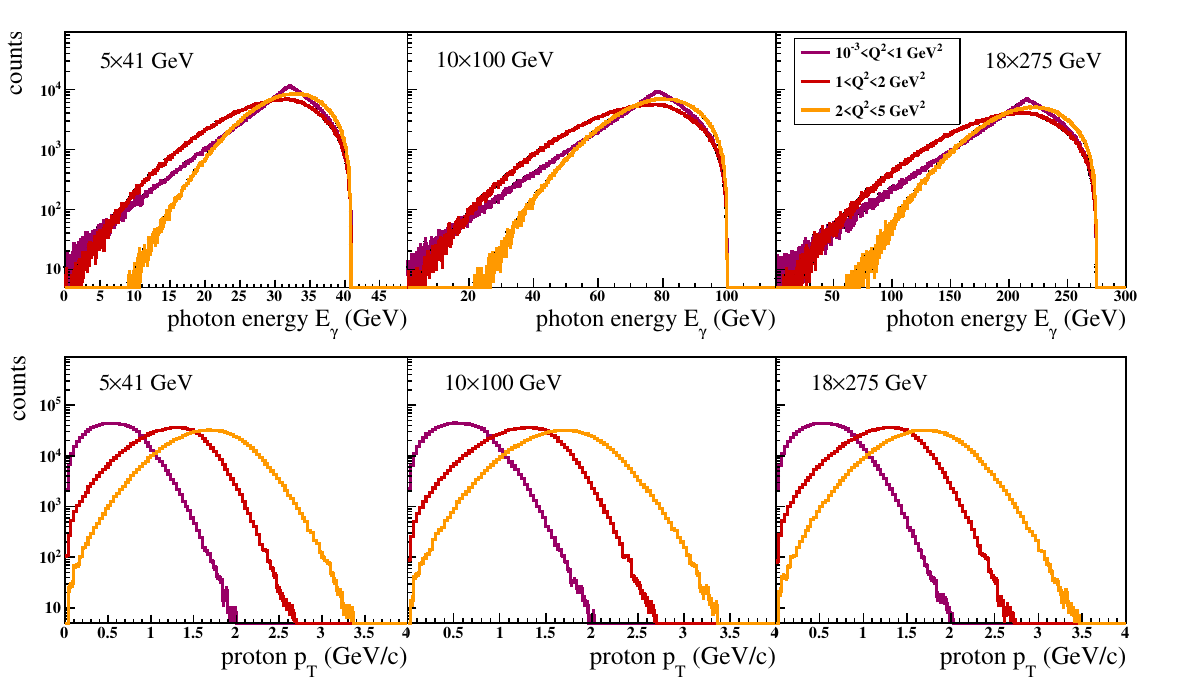}
\caption{Energy distributions of photons and $p_T$ distributions of scattered protons in backward VCS at 5$\times$41~GeV, 10$\times$100~GeV, and 18$\times$275~GeV.}
    \label{fig:protonpt}
  \end{center}
\end{figure*}

The photon rapidity is only slightly dependent on the $Q^2$ of the process. Instead, the photon rapidity is primarily dependent on collision energy; the higher the collision energy, the larger the photon rapidity. For 18$\times$275~GeV collisions, most of the photons will strike the ZDC. As the photon energy decreases, a B0 calorimeter becomes more important for photon detection. 
Geometric acceptances for simultaneous proton+photon detection are estimated in Tab.~\ref{tab:geoeff}.

The energy distribution of Compton photons is shown in Fig.~\ref{fig:protonpt} (top). Usually most of the incident proton's momentum transfers to these photons. Occasionally, however, the $\gamma^*p$ center-of-mass energy is small enough and the Mandelstam $|u|$ is large enough that the scattered photons are much lower in energy. These photons then have a large energy range, spanning from 0~GeV up to the proton beam energy. 
The $Q^2$ does not have a large effect on this energy range. This underscores one of the key takeaways from Fig.~\ref{fig:energyscan}: the final-state photon kinematics are dominated by the proton beam energy. 

Figure~\ref{fig:protonpt} (bottom) shows the $p_T$ distribution of protons at the three collision energies and $Q^2$ ranges. As expected, the proton $p_T$ distributions are almost entirely determined by the $Q^2$ of the event, and nearly independent of the collision energy.

\subsection{$\pi^0$ Background Simulations}
\label{section:simulatedbackground}

\begin{figure*}
  \begin{center}
    \includegraphics[width=0.7\textwidth]{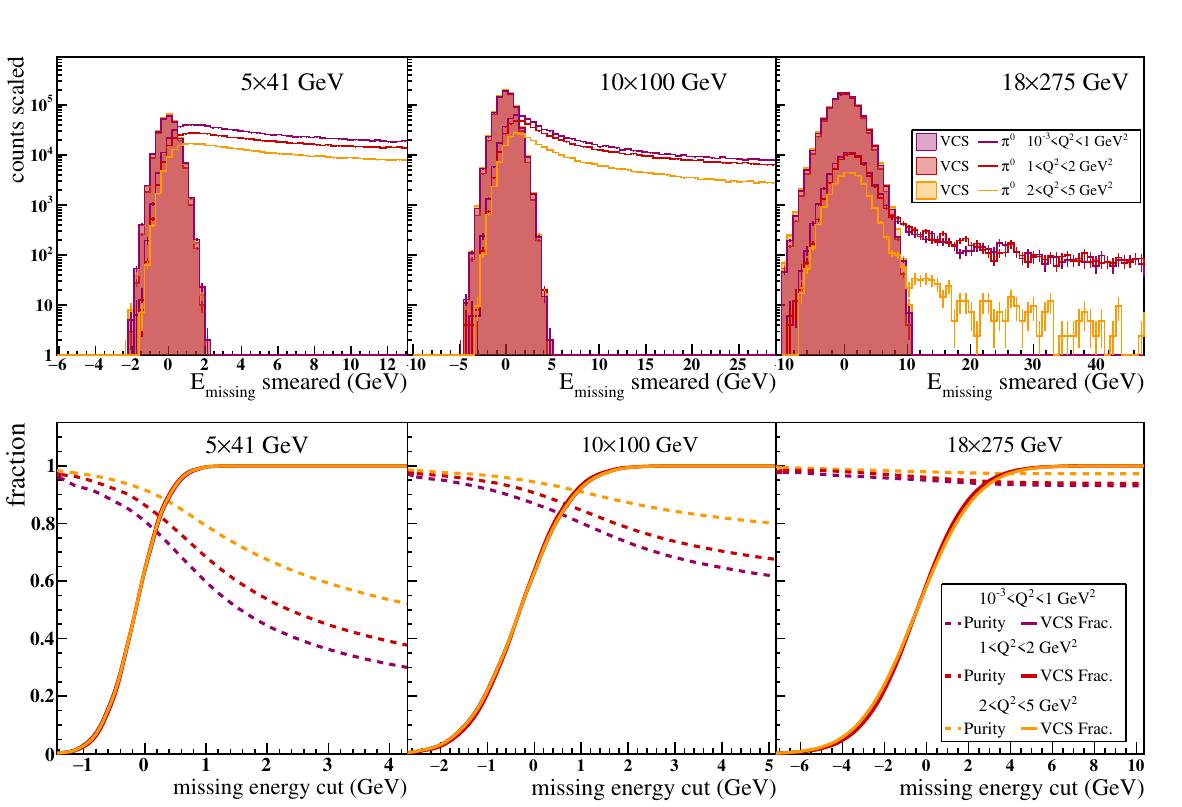}
    \caption{(top) Missing energy distribution of single photons within ZDC acceptance. The $\pi^0$ distributions are scaled to the Compton distributions by the ratio of their cross sections as shown in Tab.~\ref{tab:rates}. (bottom) Purity fraction and fraction of signal collected for a given missing energy cut.}
    \label{fig:missingenergy}
  \end{center}
\end{figure*}

To accurately measure backward Compton scattering cross sections, full event reconstruction is necessary to exclude the background from $\pi^0$ decays.
This requires three things: detecting the scattered electron, electromagnetic calorimetry in the far-forward (ion-going) region, and charged-particle tracking in the central and forward regions to detect the proton scattered toward midrapidity.

One difficulty with the $\pi^0$ background is that the two photons might merge within the same tower of the ZDC, becoming nearly indistinguishable from a single Compton photon. We may rule this out as a possibility by considering the minimum possible opening angle of the $\gamma\gamma$ pair in the lab frame:

\begin{equation}
\theta^{\gamma\gamma}_{\rm min} = 2\arctan(1/(\beta\gamma))\approx2\arctan(1/\gamma),
\end{equation}
where $\beta$ and $\gamma$ describe the Lorentz boost of the $\pi^0$ with respect to the lab frame. In backward production the $\pi^0$ is maximally boosted when it possesses almost the entire energy of the proton beam. This corresponds to the minimum possible opening angle of the photon pair. The ZDC will be $\sim$35~m downstream of the interaction point, so the minimum possible transverse separation of the photons is:
\begin{equation}
\Delta x^{\gamma\gamma}_{\text{min}} \approx (70~\text{m})m_{\pi^0}/E_{p \text{ beam}}.
\end{equation}

For proton beams at 41, 100, and 275~GeV, the minimum transverse photon separation is then 23~cm, 9.5~cm, and 3.4~cm respectively. These separations are larger than the projected 2~cm transverse widths of the ZDC towers. 
Two photons will never merge within the same tower. 

The large minimum transverse separations suggest a different issue. The dominant background will be events in which one of the $\pi^0$ photons carries the majority of the energy, and the other misses the ZDC entirely. Depending on the ZDC's resolution, the single detected photon can be mistaken for a backward Compton photon.

In order to distinguish Compton photons from $\pi^0$ photons, we require excellent high-energy resolution. Currently the ZDC is expected to have a resolution for high-energy photons~\cite{Jentsch} as:
\begin{equation}
\label{eq:zdcresolution}
\Delta E/E \sim (2\%-5\%)/\sqrt{E}\oplus1\%. 
\end{equation}
In this section, we take the 5\% upper limit for a conservative estimate. We can use exclusivity variables such as missing energy and missing $p_T$ to reduce the $\pi^0$ background. A simulation of the missing energy in VCS and $\pi^0$ events is shown in Fig.~\ref{fig:missingenergy}. In these simulations, $10^6$ $u$-channel VCS and $\pi^0$ events were simulated at each collision energy and $Q^2$ range. Only events in which a single photon landed within the ZDC's predicted 60$\times$60~cm acceptance are tabulated. 

If ePIC had perfect resolution, the single-photon $\pi^0$ events could always be rejected by their missing energy, but due to detector-resolution effects a Compton photon will not appear different from a $\pi^0$ photon. To simulate this effect, the ZDC's projected high-energy photon resolution (Eq.~\ref{eq:zdcresolution}) was used to smear these single-photon energies. Model~1, discussed in Sec.~\ref{section:themodel}, was used to generate the Compton photons, and the $\pi^0$ photon count was scaled up by the ratio of cross sections in Tab.~\ref{tab:rates}. Radiative photons may increase the measured missing energy, depending on where they go, but not enough to affect our overall conclusions.

The top panel of Fig.~\ref{fig:missingenergy} compares these missing energy distributions. At 5$\times$41~GeV, the ZDC is small enough that the $\pi^0$ photons easily miss it, and a simple cut of $E_{\rm missing}<5$~GeV reduces much of the background. However even with this cut, the background is of the same size as the Compton signal.
At 18$\times$275~GeV, the $u$-channel $\pi^0$s are so far-forward that there are very few events in which one of the photons misses the ZDC, making it easy to reject these as VCS candidates. Collisions at 18$\times$275~GeV result in $\pi^0$ photons which are highly-boosted so that they rarely miss the ZDC, but not so massively boosted that they merge within the same calorimeter tower. This makes it the optimal collider configuration for measuring both $u$-channel Compton scattering and $u$-channel $\pi^0$ production.

The bottom panel of Fig.~\ref{fig:missingenergy} demonstrates the effect of cutting on the missing energy, given a single-photon hit in the ZDC. The solid curve shows the fraction of the Compton events that would be collected given a choice of missing-energy cut: $E_{\rm missing}<E_{\rm cut}$. The dashed curve shows the purity ($P=N_{\rm VCS}/(N_{\rm VCS}+N_{\pi^0})$) of the VCS sample collected given such a cut. For example, at 5$\times$41~GeV a cut of $E_{\rm missing}<1$~GeV should be sufficient for collecting $\sim$100\% of the Compton photons; any cut larger than that would only decrease the sample purity.

A cut on missing transverse momentum ($p_T$) can also reduce the $\pi^0$ background. This was studied by smearing the $p_T$ of photons in the ZDC assuming the energy resolution discussed above and a further assumption of $2\times2$~cm$^2$ calorimeter towers. In the absence of missing-energy cuts, missing-$p_T$ cuts can improve the sample purity. However, for a missing-energy cut large enough to collect the entire VCS sample, and small enough to avoid additional $\pi^0$ events, an additional missing-$p_T$ cut does not further improve the sample purity. This conclusion is true even with improved cluster positioning resolution.

From Fig.~\ref{fig:missingenergy}, $u$-channel Compton events may be detected with 93-98\% purity at 18$\times$275~GeV with the inclusion of a cut on $E_{\rm missing}<5$~GeV. Motivated by this finding, by the acceptances tabulated in Tab.~\ref{tab:geoeff}, and the acceptance distributions shown in Fig.~\ref{fig:energyscan}, we conclude that $u$-channel Compton scattering is most easily studied in the 18$\times$275~GeV collider configuration. This conclusion is relatively model-independent, primarily due to the high photon (or $\pi^0$) energy, which leads to a high probability that the photons will hit the ZDC.

Models~1 and 2 give two extremes for backward VCS rates and rates for model~1 are around five times greater than those for model~2. Yet even for the most pessimistic rates in Tab.~\ref{tab:rates}, there is a usable number of backward VCS events at low $Q^2$. 

Overall, these models paint a promising picture for detecting $u$-channel VCS at top EIC energies, although the lower rates predicted by model~2 would make separation of the signal from the $\pi^0$ background challenging. Full detector simulations and finalized acceptances will be needed in order to further study how to reduce the background.

\section{Conclusions}
\label{section:conclusions}  
The $u$-channel contribution to the virtual Compton scattering cross section is an opportunity to measure and interpret new physics at the future EIC. Measurement of a backward peak in the VCS cross section will help clarify the mechanism by which $u$-channel production proceeds, and may shed light on which processes contribute to baryon stopping. The cross section can also be interpreted in the baryon-to-photon TDA formalism as a description of the partonic makeup of the proton in the transverse plane.

We have developed models of both $u$-channel Compton scattering and its dominant background, $u$-channel $\pi^0$ production. The Monte Carlo generator eSTARlight was used to simulate $u$-channel VCS and $\pi^0$ production at the EIC. Event rates and the distribution of final-state particles within projected detector acceptances were predicted. As the ePIC design takes shape in the coming years, more work will be needed in order to understand the effect of bremsstrahlung electrons on low-$Q^2$ event reconstruction.

In backward VCS simulations, the scattered proton lands within the acceptance of the central trackers, the B0 trackers, and the Roman pots. Protons from high-$Q^2$ events will be detected primarily in the B0 and central trackers. The Compton photon rarely ends up in the central detectors ($\eta<3.5$), and requires calorimetry in the far-forward region. The inclusion of electromagnetic calorimetry in the B0 magnet system should improve the $u$-channel VCS acceptance by a factor of two at $10\times100$~GeV and a factor of ten at $5\times41$~GeV. The additional calorimeter would also aid in the reduction of the $u$-channel $\pi^0$ background. 

In $ep$ collisions at 18$\times$275~GeV, backward VCS photons land primarily within the projected ZDC. If a $u$-channel peak has a large enough cross section ($\sim$1/10th the $\pi^0$ peak) then the $\pi^0$ background will be reducible to a few-percent level with appropriate exclusivity cuts. The feasibility of measuring backward VCS depends largely on the acceptance of the ZDC, and will be aided by the inclusion of electromagnetic calorimeters in the B0 magnet system. Assuming that the cross-sections are similar to our model predictions, $u$-channel virtual Compton scattering should be measurable at the EIC.

\section{Acknowledgments}
\label{section:acknowldegments}
We acknowledge useful conversations with Bill Li, Feng Yuan, and Alex Jentsch. Aaron Stanek participated in an early phase of this work. 
This work is supported in part by the U.S. Department of Energy, Office of Science, Office of Nuclear Physics, under contract numbers DE-AC02-05CH11231, by the US National Science Foundation under Grant No. PHY-2209614, and by the University of California Office of the President Multicampus Research Programs under grant number M21PR3502.  

\bibliographystyle{apsrev4-2} 
\bibliography{main}

\begin{thebibliography}{34}%
\makeatletter
\providecommand \@ifxundefined [1]{%
 \@ifx{#1\undefined}
}%
\providecommand \@ifnum [1]{%
 \ifnum #1\expandafter \@firstoftwo
 \else \expandafter \@secondoftwo
 \fi
}%
\providecommand \@ifx [1]{%
 \ifx #1\expandafter \@firstoftwo
 \else \expandafter \@secondoftwo
 \fi
}%
\providecommand \natexlab [1]{#1}%
\providecommand \enquote  [1]{``#1''}%
\providecommand \bibnamefont  [1]{#1}%
\providecommand \bibfnamefont [1]{#1}%
\providecommand \citenamefont [1]{#1}%
\providecommand \href@noop [0]{\@secondoftwo}%
\providecommand \href [0]{\begingroup \@sanitize@url \@href}%
\providecommand \@href[1]{\@@startlink{#1}\@@href}%
\providecommand \@@href[1]{\endgroup#1\@@endlink}%
\providecommand \@sanitize@url [0]{\catcode `\\12\catcode `\$12\catcode
  `\&12\catcode `\#12\catcode `\^12\catcode `\_12\catcode `\%12\relax}%
\providecommand \@@startlink[1]{}%
\providecommand \@@endlink[0]{}%
\providecommand \url  [0]{\begingroup\@sanitize@url \@url }%
\providecommand \@url [1]{\endgroup\@href {#1}{\urlprefix }}%
\providecommand \urlprefix  [0]{URL }%
\providecommand \Eprint [0]{\href }%
\providecommand \doibase [0]{https://doi.org/}%
\providecommand \selectlanguage [0]{\@gobble}%
\providecommand \bibinfo  [0]{\@secondoftwo}%
\providecommand \bibfield  [0]{\@secondoftwo}%
\providecommand \translation [1]{[#1]}%
\providecommand \BibitemOpen [0]{}%
\providecommand \bibitemStop [0]{}%
\providecommand \bibitemNoStop [0]{.\EOS\space}%
\providecommand \EOS [0]{\spacefactor3000\relax}%
\providecommand \BibitemShut  [1]{\csname bibitem#1\endcsname}%
\let\auto@bib@innerbib\@empty
\bibitem [{\citenamefont {Burkardt}(2003)}]{BURKARDT_2003}%
  \BibitemOpen
  \bibfield  {author} {\bibinfo {author} {\bibfnamefont {M.}~\bibnamefont
  {Burkardt}},\ }\href {https://doi.org/10.1142/s0217751x03012370} {\bibfield
  {journal} {\bibinfo  {journal} {International Journal of Modern Physics A}\
  }\textbf {\bibinfo {volume} {18}},\ \bibinfo {pages} {173} (\bibinfo {year}
  {2003})}\BibitemShut {NoStop}%
\bibitem [{\citenamefont {Akhunzyanov}\ \emph {et~al.}(2019)\citenamefont
  {Akhunzyanov} \emph {et~al.}}]{2019188}%
  \BibitemOpen
  \bibfield  {author} {\bibinfo {author} {\bibfnamefont {R.}~\bibnamefont
  {Akhunzyanov}} \emph {et~al.},\ }\href
  {https://doi.org/https://doi.org/10.1016/j.physletb.2019.04.038} {\bibfield
  {journal} {\bibinfo  {journal} {Phys. Lett. B}\ }\textbf {\bibinfo {volume}
  {793}},\ \bibinfo {pages} {188} (\bibinfo {year} {2019})}\BibitemShut
  {NoStop}%
\bibitem [{\citenamefont {Diehl}(2002)}]{Diehl:2002he}%
  \BibitemOpen
  \bibfield  {author} {\bibinfo {author} {\bibfnamefont {M.}~\bibnamefont
  {Diehl}},\ }\href {https://doi.org/10.1007/s10052-002-1016-9} {\bibfield
  {journal} {\bibinfo  {journal} {Eur. Phys. J. C}\ }\textbf {\bibinfo {volume}
  {25}},\ \bibinfo {pages} {223} (\bibinfo {year} {2002})},\ \bibinfo {note}
  {[Erratum: Eur. Phys. J. C 31, 277--278 (2003)]},\ \Eprint
  {https://arxiv.org/abs/hep-ph/0205208} {arXiv:hep-ph/0205208} \BibitemShut
  {NoStop}%
\bibitem [{\citenamefont {Defurne}\ \emph {et~al.}(2017)\citenamefont {Defurne}
  \emph {et~al.}}]{Defurne:2017paw}%
  \BibitemOpen
  \bibfield  {author} {\bibinfo {author} {\bibfnamefont {M.}~\bibnamefont
  {Defurne}} \emph {et~al.},\ }\href
  {https://doi.org/10.1038/s41467-017-01819-3} {\bibfield  {journal} {\bibinfo
  {journal} {Nature Commun.}\ }\textbf {\bibinfo {volume} {8}},\ \bibinfo
  {pages} {1408} (\bibinfo {year} {2017})},\ \Eprint
  {https://arxiv.org/abs/1703.09442} {arXiv:1703.09442 [hep-ex]} \BibitemShut
  {NoStop}%
\bibitem [{\citenamefont {Aaron}\ \emph {et~al.}(2008)\citenamefont {Aaron}
  \emph {et~al.}}]{H1:2007vrx}%
  \BibitemOpen
  \bibfield  {author} {\bibinfo {author} {\bibfnamefont {F.~D.}\ \bibnamefont
  {Aaron}} \emph {et~al.} (\bibinfo {collaboration} {H1}),\ }\href
  {https://doi.org/10.1016/j.physletb.2007.11.093} {\bibfield  {journal}
  {\bibinfo  {journal} {Phys. Lett. B}\ }\textbf {\bibinfo {volume} {659}},\
  \bibinfo {pages} {796} (\bibinfo {year} {2008})},\ \Eprint
  {https://arxiv.org/abs/0709.4114} {arXiv:0709.4114 [hep-ex]} \BibitemShut
  {NoStop}%
\bibitem [{\citenamefont {Chekanov}\ \emph {et~al.}(2003)\citenamefont
  {Chekanov} \emph {et~al.}}]{ZEUS:2003pwh}%
  \BibitemOpen
  \bibfield  {author} {\bibinfo {author} {\bibfnamefont {S.}~\bibnamefont
  {Chekanov}} \emph {et~al.} (\bibinfo {collaboration} {ZEUS}),\ }\href
  {https://doi.org/10.1016/j.physletb.2003.08.048} {\bibfield  {journal}
  {\bibinfo  {journal} {Phys. Lett. B}\ }\textbf {\bibinfo {volume} {573}},\
  \bibinfo {pages} {46} (\bibinfo {year} {2003})},\ \Eprint
  {https://arxiv.org/abs/hep-ex/0305028} {arXiv:hep-ex/0305028} \BibitemShut
  {NoStop}%
\bibitem [{\citenamefont {Accardi}\ \emph {et~al.}(2012)\citenamefont {Accardi}
  \emph {et~al.}}]{WhitePaper}%
  \BibitemOpen
  \bibfield  {author} {\bibinfo {author} {\bibfnamefont {A.}~\bibnamefont
  {Accardi}} \emph {et~al.},\ }\href
  {https://doi.org/10.1140/epja/i2016-16268-9} {\bibfield  {journal} {\bibinfo
  {journal} {Eur. Phys. J. A}\ }\textbf {\bibinfo {volume} {52}} (\bibinfo
  {year} {2012})}\BibitemShut {NoStop}%
\bibitem [{\citenamefont {Pire}\ \emph {et~al.}(2021)\citenamefont {Pire},
  \citenamefont {Semenov-Tian-Shansky},\ and\ \citenamefont
  {Szymanowski}}]{Pire:2021hbl}%
  \BibitemOpen
  \bibfield  {author} {\bibinfo {author} {\bibfnamefont {B.}~\bibnamefont
  {Pire}}, \bibinfo {author} {\bibfnamefont {K.}~\bibnamefont
  {Semenov-Tian-Shansky}},\ and\ \bibinfo {author} {\bibfnamefont
  {L.}~\bibnamefont {Szymanowski}},\ }\href
  {https://doi.org/10.1016/j.physrep.2021.09.002} {\bibfield  {journal}
  {\bibinfo  {journal} {Phys. Rept.}\ }\textbf {\bibinfo {volume} {940}},\
  \bibinfo {pages} {1} (\bibinfo {year} {2021})},\ \Eprint
  {https://arxiv.org/abs/2103.01079} {arXiv:2103.01079 [hep-ph]} \BibitemShut
  {NoStop}%
\bibitem [{\citenamefont {Cebra}\ \emph {et~al.}(2022)\citenamefont {Cebra},
  \citenamefont {Sweger}, \citenamefont {Dong}, \citenamefont {Ji},\ and\
  \citenamefont {Klein}}]{PhysRevC.106.015204}%
  \BibitemOpen
  \bibfield  {author} {\bibinfo {author} {\bibfnamefont {D.}~\bibnamefont
  {Cebra}}, \bibinfo {author} {\bibfnamefont {Z.}~\bibnamefont {Sweger}},
  \bibinfo {author} {\bibfnamefont {X.}~\bibnamefont {Dong}}, \bibinfo {author}
  {\bibfnamefont {Y.}~\bibnamefont {Ji}},\ and\ \bibinfo {author}
  {\bibfnamefont {S.~R.}\ \bibnamefont {Klein}},\ }\href
  {https://doi.org/10.1103/PhysRevC.106.015204} {\bibfield  {journal} {\bibinfo
   {journal} {Phys. Rev. C}\ }\textbf {\bibinfo {volume} {106}},\ \bibinfo
  {pages} {015204} (\bibinfo {year} {2022})}\BibitemShut {NoStop}%
\bibitem [{\citenamefont {Li}(2017)}]{billspaper}%
  \BibitemOpen
  \bibfield  {author} {\bibinfo {author} {\bibfnamefont {W.~B.}\ \bibnamefont
  {Li}},\ }\href {https://doi.org/10.2172/1408890} {\bibinfo {type} {Phd
  thesis}},\ \bibinfo  {school} {University of Regina}, \bibinfo {address}
  {Regina, Saskatchewan} (\bibinfo {year} {2017}),\ \bibinfo {note} {available
  at \url{https://www.osti.gov/biblio/1408890}}\BibitemShut {NoStop}%
\bibitem [{\citenamefont {Gayoso}\ \emph {et~al.}(2021)\citenamefont {Gayoso}
  \emph {et~al.}}]{Gayoso:2021rzj}%
  \BibitemOpen
  \bibfield  {author} {\bibinfo {author} {\bibfnamefont {C.~A.}\ \bibnamefont
  {Gayoso}} \emph {et~al.},\ }\href
  {https://doi.org/10.1140/epja/s10050-021-00625-2} {\bibfield  {journal}
  {\bibinfo  {journal} {Eur. Phys. J. A}\ }\textbf {\bibinfo {volume} {57}},\
  \bibinfo {pages} {342} (\bibinfo {year} {2021})},\ \Eprint
  {https://arxiv.org/abs/2107.06748} {arXiv:2107.06748 [hep-ph]} \BibitemShut
  {NoStop}%
\bibitem [{\citenamefont {Morand}\ \emph {et~al.}(2005)\citenamefont {Morand}
  \emph {et~al.}}]{CLAS_Morand_2005}%
  \BibitemOpen
  \bibfield  {author} {\bibinfo {author} {\bibfnamefont {L.}~\bibnamefont
  {Morand}} \emph {et~al.} (\bibinfo {collaboration} {CLAS Collaboration}),\
  }\href {https://doi.org/10.1140/epja/i2005-10032-4} {\bibfield  {journal}
  {\bibinfo  {journal} {European Physical Journal A}\ }\textbf {\bibinfo
  {volume} {24}},\ \bibinfo {pages} {445} (\bibinfo {year} {2005})}\BibitemShut
  {NoStop}%
\bibitem [{\citenamefont {Horn}\ \emph {et~al.}(2006)\citenamefont {Horn} \emph
  {et~al.}}]{PhysRevLett.97.192001}%
  \BibitemOpen
  \bibfield  {author} {\bibinfo {author} {\bibfnamefont {T.}~\bibnamefont
  {Horn}} \emph {et~al.} (\bibinfo {collaboration} {Jefferson Lab
  ${F}_{\ensuremath{\pi}}$ Collaboration}),\ }\href
  {https://doi.org/10.1103/PhysRevLett.97.192001} {\bibfield  {journal}
  {\bibinfo  {journal} {Phys. Rev. Lett.}\ }\textbf {\bibinfo {volume} {97}},\
  \bibinfo {pages} {192001} (\bibinfo {year} {2006})}\BibitemShut {NoStop}%
\bibitem [{\citenamefont {Laveissi{\`{e}}re}\ \emph {et~al.}(2009)\citenamefont
  {Laveissi{\`{e}}re} \emph {et~al.}}]{Laveissi_re_2009}%
  \BibitemOpen
  \bibfield  {author} {\bibinfo {author} {\bibfnamefont {G.}~\bibnamefont
  {Laveissi{\`{e}}re}} \emph {et~al.},\ }\bibfield  {journal} {\bibinfo
  {journal} {Phys. Rev. C}\ }\textbf {\bibinfo {volume} {79}},\ \href
  {https://doi.org/10.1103/physrevc.79.015201} {10.1103/physrevc.79.015201}
  (\bibinfo {year} {2009})\BibitemShut {NoStop}%
\bibitem [{\citenamefont {Shupe}\ \emph {et~al.}(1979)\citenamefont {Shupe}
  \emph {et~al.}}]{PhysRevD.19.1921}%
  \BibitemOpen
  \bibfield  {author} {\bibinfo {author} {\bibfnamefont {M.~A.}\ \bibnamefont
  {Shupe}} \emph {et~al.},\ }\href {https://doi.org/10.1103/PhysRevD.19.1921}
  {\bibfield  {journal} {\bibinfo  {journal} {Phys. Rev. D}\ }\textbf {\bibinfo
  {volume} {19}},\ \bibinfo {pages} {1921} (\bibinfo {year}
  {1979})}\BibitemShut {NoStop}%
\bibitem [{\citenamefont {Kumerički}\ \emph {et~al.}(2016)\citenamefont
  {Kumerički}, \citenamefont {Liuti},\ and\ \citenamefont
  {Moutarde}}]{GPDs_article}%
  \BibitemOpen
  \bibfield  {author} {\bibinfo {author} {\bibfnamefont {K.}~\bibnamefont
  {Kumerički}}, \bibinfo {author} {\bibfnamefont {S.}~\bibnamefont {Liuti}},\
  and\ \bibinfo {author} {\bibfnamefont {H.}~\bibnamefont {Moutarde}},\ }\href
  {https://doi.org/10.1140/epja/i2016-16157-3} {\bibfield  {journal} {\bibinfo
  {journal} {Eur. Phys. J. A}\ }\textbf {\bibinfo {volume} {52}} (\bibinfo
  {year} {2016})}\BibitemShut {NoStop}%
\bibitem [{\citenamefont {Budnev}\ \emph {et~al.}(1975)\citenamefont {Budnev},
  \citenamefont {Ginzburg}, \citenamefont {Meledin},\ and\ \citenamefont
  {Serbo}}]{Budnev:1975poe}%
  \BibitemOpen
  \bibfield  {author} {\bibinfo {author} {\bibfnamefont {V.~M.}\ \bibnamefont
  {Budnev}}, \bibinfo {author} {\bibfnamefont {I.~F.}\ \bibnamefont
  {Ginzburg}}, \bibinfo {author} {\bibfnamefont {G.~V.}\ \bibnamefont
  {Meledin}},\ and\ \bibinfo {author} {\bibfnamefont {V.~G.}\ \bibnamefont
  {Serbo}},\ }\href {https://doi.org/10.1016/0370-1573(75)90009-5} {\bibfield
  {journal} {\bibinfo  {journal} {Phys. Rept.}\ }\textbf {\bibinfo {volume}
  {15}},\ \bibinfo {pages} {181} (\bibinfo {year} {1975})}\BibitemShut
  {NoStop}%
\bibitem [{\citenamefont {Adloff}\ \emph {et~al.}(2000)\citenamefont {Adloff}
  \emph {et~al.}}]{H1:1999pji}%
  \BibitemOpen
  \bibfield  {author} {\bibinfo {author} {\bibfnamefont {C.}~\bibnamefont
  {Adloff}} \emph {et~al.} (\bibinfo {collaboration} {H1}),\ }\href
  {https://doi.org/10.1007/s100520050703} {\bibfield  {journal} {\bibinfo
  {journal} {Eur. Phys. J. C}\ }\textbf {\bibinfo {volume} {13}},\ \bibinfo
  {pages} {371} (\bibinfo {year} {2000})},\ \Eprint
  {https://arxiv.org/abs/hep-ex/9902019} {arXiv:hep-ex/9902019} \BibitemShut
  {NoStop}%
\bibitem [{\citenamefont {Perez}\ \emph {et~al.}(2004)\citenamefont {Perez},
  \citenamefont {Schoeffel},\ and\ \citenamefont {Favart}}]{Perez:2004ig}%
  \BibitemOpen
  \bibfield  {author} {\bibinfo {author} {\bibfnamefont {E.}~\bibnamefont
  {Perez}}, \bibinfo {author} {\bibfnamefont {L.}~\bibnamefont {Schoeffel}},\
  and\ \bibinfo {author} {\bibfnamefont {L.}~\bibnamefont {Favart}},\
  }\href@noop {} {\  (\bibinfo {year} {2004})},\ \Eprint
  {https://arxiv.org/abs/hep-ph/0411389} {arXiv:hep-ph/0411389} \BibitemShut
  {NoStop}%
\bibitem [{\citenamefont {Lomnitz}\ and\ \citenamefont
  {Klein}(2019)}]{PhysRevC.99.015203}%
  \BibitemOpen
  \bibfield  {author} {\bibinfo {author} {\bibfnamefont {M.}~\bibnamefont
  {Lomnitz}}\ and\ \bibinfo {author} {\bibfnamefont {S.}~\bibnamefont
  {Klein}},\ }\href {https://doi.org/10.1103/PhysRevC.99.015203} {\bibfield
  {journal} {\bibinfo  {journal} {Phys. Rev. C}\ }\textbf {\bibinfo {volume}
  {99}},\ \bibinfo {pages} {015203} (\bibinfo {year} {2019})}\BibitemShut
  {NoStop}%
\bibitem [{\citenamefont {Clifft}\ \emph {et~al.}(1977)\citenamefont {Clifft}
  \emph {et~al.}}]{Clifft:1977yi}%
  \BibitemOpen
  \bibfield  {author} {\bibinfo {author} {\bibfnamefont {R.~W.}\ \bibnamefont
  {Clifft}} \emph {et~al.},\ }\href
  {https://doi.org/10.1016/0370-2693(77)90082-X} {\bibfield  {journal}
  {\bibinfo  {journal} {Phys. Lett. B}\ }\textbf {\bibinfo {volume} {72}},\
  \bibinfo {pages} {144} (\bibinfo {year} {1977})}\BibitemShut {NoStop}%
\bibitem [{\citenamefont {Tompkins}\ \emph {et~al.}(1969)\citenamefont
  {Tompkins}, \citenamefont {Anderson}, \citenamefont {Gittelman},
  \citenamefont {Litt}, \citenamefont {Wiik}, \citenamefont {Yount},\ and\
  \citenamefont {Minten}}]{PhysRevLett.23.725}%
  \BibitemOpen
  \bibfield  {author} {\bibinfo {author} {\bibfnamefont {D.}~\bibnamefont
  {Tompkins}}, \bibinfo {author} {\bibfnamefont {R.}~\bibnamefont {Anderson}},
  \bibinfo {author} {\bibfnamefont {B.}~\bibnamefont {Gittelman}}, \bibinfo
  {author} {\bibfnamefont {J.}~\bibnamefont {Litt}}, \bibinfo {author}
  {\bibfnamefont {B.~H.}\ \bibnamefont {Wiik}}, \bibinfo {author}
  {\bibfnamefont {D.}~\bibnamefont {Yount}},\ and\ \bibinfo {author}
  {\bibfnamefont {A.}~\bibnamefont {Minten}},\ }\href
  {https://doi.org/10.1103/PhysRevLett.23.725} {\bibfield  {journal} {\bibinfo
  {journal} {Phys. Rev. Lett.}\ }\textbf {\bibinfo {volume} {23}},\ \bibinfo
  {pages} {725} (\bibinfo {year} {1969})}\BibitemShut {NoStop}%
\bibitem [{\citenamefont {Li}\ \emph {et~al.}(2022)\citenamefont {Li},
  \citenamefont {Stevens},\ and\ \citenamefont {Huber}}]{Li:2022dxk}%
  \BibitemOpen
  \bibfield  {author} {\bibinfo {author} {\bibfnamefont {W.~B.}\ \bibnamefont
  {Li}}, \bibinfo {author} {\bibfnamefont {J.~R.}\ \bibnamefont {Stevens}},\
  and\ \bibinfo {author} {\bibfnamefont {G.~M.}\ \bibnamefont {Huber}},\
  }\href@noop {} {\  (\bibinfo {year} {2022})},\ \Eprint
  {https://arxiv.org/abs/2205.11763} {arXiv:2205.11763 [nucl-ex]} \BibitemShut
  {NoStop}%
\bibitem [{\citenamefont {Crittenden}(1997)}]{Crittenden:1997yz}%
  \BibitemOpen
  \bibfield  {author} {\bibinfo {author} {\bibfnamefont {J.~A.}\ \bibnamefont
  {Crittenden}},\ }\href@noop {} {\  (\bibinfo {year} {1997})},\ \Eprint
  {https://arxiv.org/abs/hep-ex/9704009} {arXiv:hep-ex/9704009} \BibitemShut
  {NoStop}%
\bibitem [{\citenamefont {Klein}\ and\ \citenamefont
  {Nystrand}(1999)}]{Klein:1999qj}%
  \BibitemOpen
  \bibfield  {author} {\bibinfo {author} {\bibfnamefont {S.}~\bibnamefont
  {Klein}}\ and\ \bibinfo {author} {\bibfnamefont {J.}~\bibnamefont
  {Nystrand}},\ }\href {https://doi.org/10.1103/PhysRevC.60.014903} {\bibfield
  {journal} {\bibinfo  {journal} {Phys. Rev. C}\ }\textbf {\bibinfo {volume}
  {60}},\ \bibinfo {pages} {014903} (\bibinfo {year} {1999})},\ \Eprint
  {https://arxiv.org/abs/hep-ph/9902259} {arXiv:hep-ph/9902259} \BibitemShut
  {NoStop}%
\bibitem [{\citenamefont {Brodsky}\ \emph {et~al.}(2009)\citenamefont
  {Brodsky}, \citenamefont {Llanes-Estrada},\ and\ \citenamefont
  {Szczepaniak}}]{PhysRevD.79.033012}%
  \BibitemOpen
  \bibfield  {author} {\bibinfo {author} {\bibfnamefont {S.~J.}\ \bibnamefont
  {Brodsky}}, \bibinfo {author} {\bibfnamefont {F.~J.}\ \bibnamefont
  {Llanes-Estrada}},\ and\ \bibinfo {author} {\bibfnamefont {A.~P.}\
  \bibnamefont {Szczepaniak}},\ }\href
  {https://doi.org/10.1103/PhysRevD.79.033012} {\bibfield  {journal} {\bibinfo
  {journal} {Phys. Rev. D}\ }\textbf {\bibinfo {volume} {79}},\ \bibinfo
  {pages} {033012} (\bibinfo {year} {2009})}\BibitemShut {NoStop}%
\bibitem [{\citenamefont {Li}\ \emph {et~al.}(2019)\citenamefont {Li} \emph
  {et~al.}}]{PhysRevLett.123.182501}%
  \BibitemOpen
  \bibfield  {author} {\bibinfo {author} {\bibfnamefont {W.~B.}\ \bibnamefont
  {Li}} \emph {et~al.} (\bibinfo {collaboration} {Jefferson Lab
  ${F}_{\ensuremath{\pi}}$ Collaboration}),\ }\href
  {https://doi.org/10.1103/PhysRevLett.123.182501} {\bibfield  {journal}
  {\bibinfo  {journal} {Phys. Rev. Lett.}\ }\textbf {\bibinfo {volume} {123}},\
  \bibinfo {pages} {182501} (\bibinfo {year} {2019})}\BibitemShut {NoStop}%
\bibitem [{\citenamefont {Danagoulian}\ \emph {et~al.}(2007)\citenamefont
  {Danagoulian} \emph {et~al.}}]{PhysRevLett.98.152001_Dan}%
  \BibitemOpen
  \bibfield  {author} {\bibinfo {author} {\bibfnamefont {A.}~\bibnamefont
  {Danagoulian}} \emph {et~al.} (\bibinfo {collaboration} {Jefferson Lab Hall A
  Collaboration}),\ }\href {https://doi.org/10.1103/PhysRevLett.98.152001}
  {\bibfield  {journal} {\bibinfo  {journal} {Phys. Rev. Lett.}\ }\textbf
  {\bibinfo {volume} {98}},\ \bibinfo {pages} {152001} (\bibinfo {year}
  {2007})}\BibitemShut {NoStop}%
\bibitem [{\citenamefont {Lansberg}\ \emph {et~al.}(2007)\citenamefont
  {Lansberg}, \citenamefont {Pire},\ and\ \citenamefont
  {Szymanowski}}]{LANSBERG200716}%
  \BibitemOpen
  \bibfield  {author} {\bibinfo {author} {\bibfnamefont {J.}~\bibnamefont
  {Lansberg}}, \bibinfo {author} {\bibfnamefont {B.}~\bibnamefont {Pire}},\
  and\ \bibinfo {author} {\bibfnamefont {L.}~\bibnamefont {Szymanowski}},\
  }\href {https://doi.org/https://doi.org/10.1016/j.nuclphysa.2006.10.014}
  {\bibfield  {journal} {\bibinfo  {journal} {Nuclear Physics A}\ }\textbf
  {\bibinfo {volume} {782}},\ \bibinfo {pages} {16} (\bibinfo {year} {2007})},\
  \bibinfo {note} {proceedings of the 5th International Conference on
  Perspectives in Hadron Physics, Particle–Nucleus and Nucleus–Nucleus
  Scattering at Relativistic Energies}\BibitemShut {NoStop}%
\bibitem [{\citenamefont {Laveissi{\`{e}}re}\ \emph {et~al.}(2004)\citenamefont
  {Laveissi{\`{e}}re} \emph {et~al.}}]{PhysRevC.69.045203}%
  \BibitemOpen
  \bibfield  {author} {\bibinfo {author} {\bibfnamefont {G.}~\bibnamefont
  {Laveissi{\`{e}}re}} \emph {et~al.} (\bibinfo {collaboration} {Jefferson Lab
  Hall A Collaboration}),\ }\href {https://doi.org/10.1103/PhysRevC.69.045203}
  {\bibfield  {journal} {\bibinfo  {journal} {Phys. Rev. C}\ }\textbf {\bibinfo
  {volume} {69}},\ \bibinfo {pages} {045203} (\bibinfo {year}
  {2004})}\BibitemShut {NoStop}%
\bibitem [{\citenamefont {Adkins}\ \emph {et~al.}(2022)\citenamefont {Adkins}
  \emph {et~al.}}]{Adkins:2022jfp}%
  \BibitemOpen
  \bibfield  {author} {\bibinfo {author} {\bibfnamefont {J.~K.}\ \bibnamefont
  {Adkins}} \emph {et~al.} (\bibinfo {collaboration} {The ECCE
  Collaboration}),\ }\href@noop {} {\  (\bibinfo {year} {2022})},\ \Eprint
  {https://arxiv.org/abs/2209.02580} {arXiv:2209.02580 [physics.ins-det]}
  \BibitemShut {NoStop}%
\bibitem [{\citenamefont {Adam}\ \emph {et~al.}(2022)\citenamefont {Adam} \emph
  {et~al.}}]{Adam_2022}%
  \BibitemOpen
  \bibfield  {author} {\bibinfo {author} {\bibfnamefont {J.}~\bibnamefont
  {Adam}} \emph {et~al.} (\bibinfo {collaboration} {The ATHENA
  Collaboration}),\ }\href {https://doi.org/10.1088/1748-0221/17/10/P10019}
  {\bibfield  {journal} {\bibinfo  {journal} {Journal of Instrumentation}\
  }\textbf {\bibinfo {volume} {17}}\bibinfo  {number} { (10)},\ \bibinfo
  {pages} {P10019}}\BibitemShut {NoStop}%
\bibitem [{\citenamefont {{Abdul Khalek}}\ \emph {et~al.}(2022)\citenamefont
  {{Abdul Khalek}} \emph {et~al.}}]{ABDULKHALEK2022122447}%
  \BibitemOpen
\bibfield  {number} {  }\bibfield  {author} {\bibinfo {author} {\bibfnamefont
  {R.}~\bibnamefont {{Abdul Khalek}}} \emph {et~al.},\ }\href
  {https://doi.org/https://doi.org/10.1016/j.nuclphysa.2022.122447} {\bibfield
  {journal} {\bibinfo  {journal} {Nuclear Physics A}\ }\textbf {\bibinfo
  {volume} {1026}},\ \bibinfo {pages} {122447} (\bibinfo {year}
  {2022})}\BibitemShut {NoStop}%
\bibitem [{\citenamefont {Jentsch}()}]{Jentsch}%
  \BibitemOpen
  \bibfield  {author} {\bibinfo {author} {\bibfnamefont {A.}~\bibnamefont
  {Jentsch}},\ }\href@noop {} {}\bibinfo {howpublished} {personal
  communication, (2023)}\BibitemShut {NoStop}%
\end{thebibliography}%
\end{document}